\documentclass[a4paper,12pt]{article}

\usepackage{amssymb,amsmath,latexsym}
\usepackage{graphicx}
\usepackage{psfrag}
\usepackage{mathtext}
\usepackage{longtable}
\usepackage{hyperref}

\newcommand{\be}{\begin{equation}}
\newcommand{\ee}{\end{equation}}
\newcommand{\beqa}{\begin{eqnarray}}
\newcommand{\eeqa}{\end{eqnarray}}

\newcommand{\pd}{\partial}
\newcommand\m{\mu}
\newcommand\D{\Delta}
\newcommand\n{\nu}
\newcommand\s{\sigma}

\newcommand\vk{\varkappa}

\def\const{\mbox{const}}
\def\e{{\rm e}}
\newcommand{\bseq}{\begin{subequations}}
\newcommand{\eseq}{\end{subequations}}

\renewcommand{\ln}{\mathop{\rm ln}\nolimits}

\newcommand{\di}{\mathrm d}

\title{
\sc{\Huge Testing
Lorentz invariance of\\ dark matter}
\date{}
\author{
  Diego Blas$^{a}$\footnote{diego.blas@cern.ch}, 
Mikhail M. Ivanov$^{b,c}$\footnote{mm.ivanov@physics.msu.ru}~ 
and Sergey Sibiryakov$^{c,b}$\footnote{sibir@inr.ac.ru; sergey.sibiryakov@cern.ch}
\vspace{.2cm}\\
\normalsize\llap{$^a$}\it Theory Group, Physics Department, CERN,
CH-1211 Geneva 23, Switzerland  \\
\normalsize\llap{$^b$}\it Faculty of Physics, Moscow State University,
\normalsize \it Vorobjevy Gory, 119991 Moscow, Russia\\
\normalsize\llap{$^c$}\it
Institute for Nuclear Research of the
Russian Academy of Sciences, \\ 
\normalsize \it  60th October Anniversary Prospect, 7a, 117312
Moscow, Russia}
}

\begin{document}

\maketitle

\vspace{-10cm}
\begin{flushright}  
CERN-PH-TH/2012-234
\end{flushright}
\vspace{8cm}
\begin{abstract}
We study the possibility to constrain 
deviations from Lorentz invariance in dark
matter (DM) with cosmological observations. 
Breaking of 
Lorentz invariance generically introduces new light
gravitational degrees of freedom, which we represent 
through a dynamical timelike vector field.
If DM does not obey Lorentz invariance, it couples to this
vector field. We find that this coupling affects the inertial mass of
small DM halos which 
no longer satisfy the equivalence principle. For large enough lumps of
DM we identify 
a (chameleon) mechanism that restores the inertial mass to its
standard value. 
As a consequence,
the dynamics of gravitational 
clustering are modified. Two prominent effects are a 
scale dependent enhancement in the growth of large scale structure 
and a scale dependent bias
between DM and baryon density perturbations.  
The comparison with the measured 
linear matter power spectrum in principle allows to bound the departure from
Lorentz invariance of DM at the 
per cent level.
\end{abstract}

\section{Introduction}

Elucidating the nature of dark matter (DM) stands as a major challenge
of modern cosmology and particle physics. One of the basic properties
always assumed is that DM
satisfies Lorentz invariance (LI). In this work we want to
analyze how this assumption can be verified from 
the study of cosmological perturbations.  

Lorentz invariance is one of the best tested symmetries of the Standard Model
of particle physics
\cite{Kostelecky:2008ts}. 
Thus, it is  tempting to postulate that it is a fundamental property
of all fields of Nature including gravity and the dark sectors of the
Universe, i.e DM and dark energy.
This 
assumption is very powerful, but also restrictive. When applied to  gravitation
 it has 
 far-reaching conceptual implications, like
 the uniqueness of General Relativity (GR) as the LI theory of
 gravity, see
 e.g. \cite{Weinberg:1965rz,Deser:1969wk,Wald:1986bj,Blas:2007pp}.
However, beautiful as it is, GR suffers from the problem of
non-renormalizability precluding its interpretation as a UV complete
quantum theory. This is the essence of the notorious problem of
quantum gravity.
 
It is conceivable that the eventual theory of quantum gravity will involve
violation of LI (Lorentz violation, LV, for short) in some form. 
For instance, it
was recently suggested by 
P.~Ho\v rava that a UV completion of GR may be possible within
perturbative quantum field theory
 at the cost of abandoning LI at very high energies
 \cite{Horava:2009uw}. This approach appears promising as it
 allows for an immediate improvement compared to GR coming from 
the softening of the gravitational
 amplitudes in the UV\footnote{One should keep in mind though that the
   renormalizability of gravity in the strict sense along these lines
   has not yet been demonstrated due to the complexity of the
   resulting theory.} \cite{Blas:2009ck}. 
If deviations from LI are present in quantum gravity, it is mandatory
to understand which are the consequences for the rest of fields in Nature \cite{Mattingly:2005re,Jacobson:2005bg}.

A first relevant observation is that even when introduced at very high
energies, LV has also 
consequences at 
low energies \cite{Blas:2009yd}. 
Indeed, LV corresponds to the existence
of a preferred reference frame at every point of
space-time. Additional local fields   
are required
to set up this frame; those have typically  massless excitations and 
affect physics at any scale.
The situation at low-energies is encapsulated by the
so-called Einstein-aether model \cite{Jacobson:2000xp,Jacobson:2008aj}
where the preferred frame is determined by a vector field with unit norm
(aether). This model is an effective theory
with a cutoff that can be as high as
the Planck mass \cite{Withers:2009qg}.
Restricting the
aether to its longitudinal component one obtains the khronometric
model that was introduced in \cite{Blas:2010hb} as the low-energy limit
of the Ho\v rava's 
proposal we just mentioned \cite{Blas:2009ck,Jacobson:2010mx} 
(more precisely, of its
healthy extended version \cite{Blas:2009qj}).
In both Einstein-aether and khronometric cases, the interaction of the aether with the metric field
has consequences for gravitation at all distances, which leads to
observable
astrophysical and cosmological effects 
\cite{Jacobson:2008aj,Mukohyama:2010xz,Blas:2011zd}.

Precision tests of LI in the Standard
Model 
of particle physics put extremely tight constraints on the coupling of
the aether to 
ordinary matter \cite{Kostelecky:2008ts,Mattingly:2005re}. If similar bounds held for the couplings of the aether
to other sectors of the Universe, LV would have no effect on
cosmology.
In this paper we adopt the viewpoint that 
there is a mechanism that enforces LI of the Standard Model with the
required precision while 
allowing for sizable LV in gravitational physics, 
DM or dark energy. Finding such mechanism while avoiding fine-tuning
is an important
challenge for the model and a few possibilities have been proposed in
\cite{GrootNibbelink:2004za,Pujolas:2011sk,Pospelov:2010mp}. 
Some consequences of LV for dark energy were
unveiled in \cite{Blas:2011en}, where it was shown that the presence of
the aether vector field 
allows to attribute the current 
acceleration of the Universe to a renormalizable
operator not sensitive to UV corrections.
The present work is devoted to a systematic investigation of LV
effects in the cosmological evolution of DM. To isolate these effects
we will assume throughout the paper that 
the dark energy is represented by the cosmological constant. 

We consider a general set-up, with DM
described in terms of an effective fluid that interacts with
the aether to account for LV. 
The simplest DM candidate described within our formalism is
the theory of massive non-relativistic particles, but other
possibilities, such as 
 axionic DM \cite{Duffy:2009ig}, are also covered.
We will assume that the velocities of DM are non-relativistic
during the observable evolution of the Universe. 
One might think that this would suppress all LV effects. However, this
is not the case. We will see that 
the coupling to the
aether modifies the inertial mass of the DM particles, but does not
affect their gravitational mass. 
As a consequence, DM violates the
equivalence
principle and the dynamics of 
cosmological perturbations change. The situation
resembles that in LI theories with additional long-range interaction
in the DM sector  
\cite{FERMILAB-PUB-91-077-A,FERMILAB-PUB-92-008-A}. 
However, the specific
signatures of LV allow to distinguish between these two cases. 

It is worth pointing out that a future direct detection of DM may
yield strong constraints on LV in this sector. Indeed, a direct detection
would imply a relatively strong 
coupling between the DM and visible sectors.
In that case LV would be transferred from DM to 
ordinary matter via loop diagrams and thus would be subject to the tight
existing bounds on LV in the Standard Model. Similar arguments were used in
\cite{Bovy:2008gh,Carroll:2008ub} to constrain violation of the
equivalence principle in the DM sector in the LI context. However, in
the absence of any direct detection so far and since 
the loop calculations are rather model-dependent, we find it
useful to study the bounds on LV in DM following from its
gravitational manifestations, in particular in cosmology.

Our work is organized as follows. In Sec.~\ref{sec:aether} we
briefly review the Einstein-aether and khronometric models. To
understand the qualitative effects of LV in DM we study in
Sec.~\ref{sec:Newton} the dynamics of massive point particles coupled
to the aether, concentrating on the Newtonian limit and Jeans
instability. We next proceed 
to the systematic treatment of LV DM using the effective
fluid description. Sec.~\ref{sec:fluid} contains the
description of the setup, analysis of the homogeneous cosmology and
the equations for the linearized perturbations.
Sec.~\ref{sec:CosmoPert} contains the
analytic study of the perturbations in various regimes and the
qualitative discussion of the observational signatures. The results of
the numerical integration of the linearized equations in a simplified
cosmological model are presented in  
Sec.~\ref{sec:numerics}. Section~\ref{sec:conclusions} contains the
summary and discussion of our results. 
Some technical details of our numerical procedure are described in 
the Appendix.

\section{Gravity with a preferred frame}
\label{sec:aether}

To describe LV we assume that at every point of space-time there is a
time-like vector $u_\m$ that following Ref.~\cite{Jacobson:2000xp} we will
call ``aether''. The vector is constrained
to have unit norm\footnote{We use the metric with $(+,-,-,-)$
  signature. Latin indices from the middle of the alphabet take 
the values $i,j=1,2,3$, while Greek
  letters denote the space-time indices. The latter 
are manipulated with $g_{\m\n}$.
  Objects in bold face are three-vectors. We
  use units where the speed of propagation of light is $c=1$.},
\be
\label{unitnorm}
u_\m u^\m=1\;.
\ee  
Thus it does not vanish anywhere and sets the preferred time-direction
at every point of  space-time. 
This breaks the local Lorentz group $SO(3,1)$ of GR
 to the local subgroup $SO(3)$ of purely spatial rotations that leave
$u_\m$ invariant.
 This way, the introduction of $u_\m$ allows us
to describe LV effects with an action invariant under
arbitrary coordinate transformations. We will be interested in the
dynamics of $u_\mu$ at large distances, which according to the rules of 
effective field theory is governed by operators
with the smallest number of derivatives. It is straightforward to see
that it is impossible to construct any contribution to the Lagrangian 
with one or
no derivatives. Thus LV at low-energies is governed by the action of
the Einstein-aether 
model \cite{Jacobson:2000xp,Jacobson:2008aj}, the most general action
containing up to two derivatives of $u_\m$,
\be
\label{Saether}
S_{\text{\ae}}\equiv-\frac{M_0^2}{2}\int \di^4x\sqrt{-g} 
\Big[R+K^{\m\n}_{~~~\s\rho}\nabla_\m u^\s\nabla_\n u^\rho+l(u_\m u^\m-1)\Big]\;,
\ee
where
\be
\label{Kmnsr}
K^{\m\n}_{~~~\s\rho}\equiv c_1g^{\m\n}g_{\s\rho}+c_2\delta_\s^\m\delta_\rho^\n
+c_3\delta_\rho^\m\delta_\s^\n+c_4u^\m u^\n g_{\s\rho},
\ee
and the last term with the Lagrange multiplier $l$ has been added to
enforce 
the unit-norm
constraint. We have included in the above action the Einstein-Hilbert
term for the metric $g_{\m\n}$. The parameter $M_0$ is
related to the Planck mass, cf. (\ref{GNaether}), while the dimensionless constants $c_a$,
$a=1,2,3,4$, 
characterize the strength of the interaction of the aether  with gravity. As
discussed below, observations require these constants to be much
less than 1, so we will assume
\be
\label{smallci}
|c_a|\ll 1\;
\ee
throughout the paper. 
Let us stress again that while we are interested
in describing LV, the action (\ref{Saether}) is explicitly generally
covariant. This stems from the assumption that the non-vanishing
aether field represents the only source of LV. In a certain sense, the picture is analogous to the
spontaneous symmetry breaking of internal symmetries\footnote{There
  is an important difference though. The condition (\ref{unitnorm})
  excludes from consideration any LI states thus precluding the
  symmetry restoration. The attempt to make the length of
  $u_\m$ dynamical to allow for the restored symmetry phase leads to
  appearance of a ghost degree of freedom that spoils the
  theory \cite{Elliott:2005va}. The absence of a healthy LI phase is
  common to gravitational models with LV, which raises doubts about
  the possibility to UV complete them within a LI setting, see the
  discussion of this issue in \cite{Dubovsky:2005xd,Adams:2006sv}.}.
  Finally, as an effective theory 
Einstein-aether has a cutoff of order $M_0|c_a|^{1/2}$
\cite{Withers:2009qg}. 
If $c_a$ are
not extremely small this
cutoff is only a
few orders of magnitude below the Planck mass.

A variant of the Einstein-aether model is obtained by
restricting the aether to be hy\-per\-sur\-face-orthogonal,
\be
\label{aetherkhron}
u_\m\equiv\frac{\pd_\m\s}{\sqrt{g^{\m\n}\pd_\m\s\pd_\n\s}}\;.
\ee
In this case the unit-norm constraint is identically satisfied and the
Lagrange multiplier term in the action (\ref{Saether}) can be omitted.
The expression (\ref{aetherkhron}) is invariant under reparameterizations
\be
\label{reparam}
\s\mapsto\tilde\s(\s)\;,
\ee
where $\tilde\s(\s)$ is an arbitrary monotonic function.
The scalar field $\s$ is assumed to have a time-like gradient 
at every point of space-time. This
 defines a preferred time-coordinate, hence $\sigma$ is called
``khronon'' (from the Greek word for
``time''), and the class of models including the metric and 
$\sigma$ ``khronometric'' models. 
When restricted to the form
 (\ref{aetherkhron}), the curl of the aether vanishes
\be
\label{vorticity}
\omega^\m\equiv\epsilon^{\m\n\lambda\rho}u_\n\nabla_\lambda u_\rho=0\;.
\ee
As a consequence, the four combinations with derivatives of $u_\m$ in
the action (\ref{Saether}) are not independent.
It is common to eliminate the
$c_1$-term in favor of the other three. The coefficients for the last
three operators in (\ref{Kmnsr}) then become
\be
\label{khpar}
\lambda\equiv c_2~,~~~\beta\equiv  c_3+c_1~,~~~~
\alpha\equiv c_4+c_1\;.
\ee
This model naturally arises as the low-energy limit
of  Ho\v rava gravity \cite{Jacobson:2010mx,Blas:2010hb}. 
In other words, Ho\v rava gravity can provide a UV completion for
the khronometric theories which potentially improves their UV behavior as compared
to GR. 
 
The main difference between Einstein-aether and khronometric models
is the number of degrees of freedom. 
Einstein-aether describes three
types of massless propagating modes: the standard transverse-traceless
tensor modes of the metric and the vector and scalar polarizations
coming from the transverse and longitudinal fluctuations of the aether.
For the khronometric case, the transverse vector polarization is absent.
In
the scalar and tensor sectors the two models are almost
equivalent\footnote{\label{footnote_ins} The only difference is the
  presence of an instantaneous mode 
in the khronometric model, 
which is absent in Einstein-aether \cite{Blas:2011ni}. This difference
is not relevant for the local physics that we study in the present
work.}. 
As in this paper we are mostly interested in
the scalar cosmological perturbations, we will often use without loss
of generality the terminology of the khronometric model and, in
particular, the constants (\ref{khpar}).
One can show
that the parameters $c_a$ 
(or \eqref{khpar} for the khronometric case) can be chosen such that all
modes are stable and have positive energy \cite{Jacobson:2008aj,Blas:2009qj}.
To avoid gravitational
and aether Cherenkov losses by high-energy cosmic rays
\cite{Elliott:2005va}, one should also  require that the velocities 
of
the graviton and aether modes are not less than the speed of
high-energy particles comprising the cosmic rays, which is close to one.  

The phenomenology of theories with action (\ref{Saether}) has 
been extensively studied  \cite{Jacobson:2008aj,Blas:2010hb,Blas:2011zd}. 
The high precision with which LI is tested within the Standard Model
excludes the direct interaction of the aether with baryonic
matter\footnote{Examples of mechanisms to suppress such interaction
  are discussed in
  \cite{GrootNibbelink:2004za,Pujolas:2011sk,Pospelov:2010mp}.},  
meaning that the latter couples universally to the metric $g_{\m\nu}$.
Still, the aether affects the gravitational field of matter
sources. 
 At the Newtonian level the modifications amount 
 to an  unobservable  renormalization of the gravitational constant that now
reads \cite{Carroll:2004ai,Blas:2009qj},
\be
\label{GNaether}
G_{N}\equiv\frac{1}{8\pi M_0^2}\bigg(1-\frac{c_1+c_4}{2}\bigg)^{-1}\;.
\ee
The deviations from GR for Solar System physics are encoded in
 two post-Newtonian parameters
$\alpha_1^{PPN}$ and $\alpha_2^{PPN}$. These have been calculated for
the generic aether in Ref.~\cite{Foster:2005dk} and for the
khronometric model in Refs.~\cite{Blas:2010hb,Blas:2011zd}.
Observations 
yield the constraints \cite{Will:2005va}
\be
\label{al12bounds}
|\alpha_1^{PPN}|\lesssim 10^{-4}~,~~~~~
|\alpha_2^{PPN}|\lesssim 4\cdot 10^{-7}\;.
\ee
Those constraints are trivially satisfied in GR, where $\alpha_1^{PPN}=\alpha_2^{PPN}=0$.
For LV theories with generic parameters 
they
imply the condition
\be
\label{firstb}
|c_a|\lesssim 10^{-7}\;.
\ee
This bound is relaxed for certain relations between the
parameters. In the generic aether model one can impose two
restrictions on $c_a$ to make both 
$\alpha_1^{PPN}$ and $\alpha_2^{PPN}$ vanish \cite{Jacobson:2008aj}.
Then one is left with a
two-parameter family of theories that are
indistinguishable from GR at the post-Newtonian
level (i.e. for Solar System tests \cite{Will:Book}). Remarkably, for
the khronometric case the same is 
achieved by the  single condition $\alpha=2\beta$, which leaves the
parameters $\beta$ and $\lambda$ 
arbitrary. Further bounds of order 
\be
\label{thirdb}
|c_a|\lesssim 10^{-2}\;
\ee
 follow from considerations of  Big Bang Nucleosynthesis
(BBN) \cite{Carroll:2004ai} and emission of gravitational waves by binary systems
\cite{Foster:2006az,Blas:2011zd}. 

To sum up, the theory (\ref{Saether}) with universal coupling of
ordinary matter to the
metric $g_{\mu\nu}$ is
phenomenologically
viable provided the stability constraints
are satisfied and \eqref{smallci} holds at the level of
\eqref{firstb} or \eqref{thirdb} (in the latter case the relations
ensuring vanishing of the PPN parameters must be fulfilled).

\section{Lorentz violating dark matter: point-particles}\label{sec:Newton}

To grasp the physical consequences of LV in the DM sector we
first study how the coupling to the aether affects the dynamics of
gravitationally interacting point
particles. This will allow us to understand the Newtonian limit of
the theory and the effects of LV on the Jeans
instability. 

In the presence of the aether the relativistic action for a massive
point particle
can be generalized to
\be
\label{1paether1}
 S_{pp}\equiv-m\int \di s\; F(u_\m v^\m)\;,
\ee 
where 
\be
\di s\equiv\sqrt{g_{\m\n}\di x^\m \di x^\n}
\ee
is the proper length along the trajectory of the particle and
\be
\label{4vel}
v^\m\equiv\frac{\di x^\m}{\di s}
\ee
is the particle's four-velocity. $F$ is an arbitrary positive function
that we normalize to $F(1)=1$; the GR limit corresponds to $F\equiv
1$. Note that particles described by the action 
(\ref{1paether1}) violate
the equivalence principle and, actually, do not follow geodesics
of any metric.

It is instructive to work out the relation between the covariant description
based on the action (\ref{1paether1}) and
the approach commonly adopted in the study of LV in microphysics. In
the latter approach 
LV manifests itself for free particles in the
modified dispersion relation \cite{Liberati:2012th}
\be
\label{dispp}
E^2={\cal E}^2({\bf p}^2)\;,
\ee
where $E$ and ${\bf p}$ are the energy and momentum of the particle in
the preferred frame where the aether is aligned with the time
direction, $u_0=1,~u_i=0$, and ${\cal E}^2$ is a function that takes
the form, 
\[
{\cal E}^2=m^2+{\bf p}^2\;,
\] 
if LI is preserved. 
The preferred frame singles out a preferred time coordinate $t$
and through (\ref{1paether1}) defines the Lagrangian
\be
\label{Lagrant}
S_{pp}=\int \di t\;  L_{pp}\;,
\ee
where
\be
L_{pp}=-m\sqrt{1-{\bf V}^2}\;F\bigg(\frac{1}{\sqrt{1-{\bf
      V}^2}}\bigg)~,~~~~~
V^i\equiv \frac{\di x^i}{\di t}\;.
\ee
The energy and momentum are now related to the velocity $V^i$ by the
standard formulas
\be
E=V^i\;\frac{\partial L_{pp}}{\partial V^i}- L_{pp}, \quad
p_i=\frac{\partial L_{pp}}{\partial V^i}\;. 
\ee
For a given function $F$ these equations yield an implicit relation
between $E$ and $p_i$, i.e. the dispersion relation of the form
(\ref{dispp}). Vice versa, starting from the dispersion relation
(\ref{dispp}) one obtains the equation for the Lagrangian
\be
\label{equ}
\left(V^i\;\frac{\partial  L_{pp}}{\partial V^i}-L_{pp}\right)^2
={\cal E}^2\bigg(\frac{\pd L_{pp}}{\pd V^i}\frac{\pd L_{pp}}{\pd V^i}\bigg)\;.
\ee
We conclude that there is a one-to-one correspondence between the
functions $F$ of the covariant approach and the functions ${\cal E}^2$
appearing in the free-particle dispersion relation.

Let us solve Eq.~(\ref{equ}) in the simplest case of the quadratic
dispersion,
\be
\label{quadrdisp}
E^2=m^2+(1+\xi){\bf p}^2\;,
\ee
where $\xi$ is a dimensionless constant. Under fairly general
assumptions this type of dispersion relations with $\xi\neq 0$
describes the leading effect of LV at relatively low energies
\cite{Coleman:1998ti}. In this case we observe that by the redefinition
$\tilde V^i\equiv V^i (1+\xi)^{-1/2}$ equation (\ref{equ}) reduces 
to the standard
relativistic form. Thus we obtain,
\be
L_{pp}=-m\sqrt{1-{\bf \tilde V}^2}=-m\sqrt{1-\frac{{\bf V}^2}{1+\xi}}\;,
\ee
and finally
\be
\label{actpp}
 S_{pp}=
-m\int \di s \sqrt{\frac{1+\xi\left(u_\mu v^\mu\right)^2}{1+\xi}}\;.
\ee
As discussed above,
this case encompasses a wide class of physically interesting
situations. However, since imposing the specific form (\ref{actpp}) does
not 
simplify the analysis, and also to be completely general, we will
continue to work with an arbitrary function~$F$. 

\subsection{Newtonian limit}
\label{ssec:newt}

We will assume that DM is non-relativistic during the relevant stages of 
cosmological evolution, i.e. it moves slowly with respect to the
cosmic frame. The latter is defined as the frame where the cosmic
microwave background (CMB) is approximately isotropic. We will assume
that it coincides with the preferred frame set by the background value
of the aether\footnote{This is justified since 
it has been shown \cite{aetheralignment} that in a
  homogeneous expanding universe the aether field tends to align with
  the time direction.}. Thus one can use the Newtonian
limit to describe the
dynamics at subhorizon scales. This corresponds to expanding the action (\ref{1paether1}) to
 quadratic order in the particle three-velocities $V^i$,
  quadratic order in the spatial component of the 
aether\footnote{We assume that $u^i$ is of the same order as $V^i$, as
will be verified by the calculation for the self-gravitating situation given below. 
The presence of the direct
coupling between DM particles and the aether is crucial for that. In the absence of such coupling,
as it happens for the field produced by 
visible matter \cite{Foster:2005dk,Blas:2011zd}, the aether
has the order $O(V^3)$ and the dynamics of the system is different.}
$u^i$ and to  linear order in the Newton potential $\phi$. 
The latter
appears in the standard Newtonian limit of the metric\footnote{
Other metric perturbations are of higher post-Newtonian order.},
\be
\label{gNewtonian}
g_{00}=1+2\phi~,~~~ g_{0i}=0~,~~~g_{ij}=-\delta_{ij}(1-2\psi)\;,
\ee
where for the moment we have neglected the cosmological expansion.
We obtain
\be
\label{1pNew}
 S_{pp}=m\int \di t\; \bigg[\frac{(V^i)^2}{2}-\phi
-Y\frac{(u^i-V^i)^2}{2}\bigg]\;,
\ee
where we have denoted
\be
\label{Ys}
Y\equiv F'(1)
\ee
and omitted the constant term corresponding to the rest-mass. To
understand the effect of LV, we consider first the case when the aether
fluctuations are negligible, $u^i=0$. In this case the last term in (\ref{1pNew})
renormalizes the particle's inertial mass
\[
m\mapsto m(1-Y)\;.
\]
On the other hand, the gravitational mass (the source of $\phi$ in
\eqref{1pNew}) remains 
equal to $m$, which clearly violates the equivalence principle.
To guarantee the positivity of the kinetic energy we impose the
restrictions $m>0$ and  $Y<1$.

Let us now consider the generic situation in the Newtonian limit for a 
dense medium composed
of DM particles. It is convenient to introduce the mass
density,
\be
\label{rhopp}
\rho({\bf x},t)=m\sum_{A}\delta^{(3)}({\bf x}-{\bf x}_A(t))\;,
\ee 
where ${\bf x}_A(t)$ is the trajectory of the $A$-th particle, and
rewrite the action (\ref{1pNew}) in the form,
\be
\label{fluidNew}
 S_{pp}=\int \di^4x\; \rho\;\bigg[\frac{(V^i)^2}{2}-\phi
-Y\frac{(u^i-V^i)^2}{2}\bigg]\;.
\ee
One observes that inside the medium the aether perturbations acquire a
quadratic potential with the central value set by the velocity of the
medium.  Not to destabilize the aether the potential must be
positive\footnote{The case of negative $Y$ can also be interesting from
the phenomenological viewpoint if the growth of the instability induced
by the negative potential is smaller or comparable to the Hubble rate.
To satisfy this requirement the absolute value of $Y$ must be
of order $c_a$, cf. Eq.~(\ref{meffh}), which already sets a strong
restriction on LV in DM.},
$Y>0$.  
We can anticipate that due to this potential the aether  tends to
align with the velocity of the medium. When alignment occurs, 
 the last term in
(\ref{fluidNew})  disappears, restoring the standard action for the
fluid universally coupled to gravity. In other words, the violation of
equivalence
principle will be screened inside a dense medium, realizing
an analog of the chameleon mechanism \cite{Khoury:2003aq}. This
picture is elaborated quantitatively in what follows.

The action (\ref{fluidNew}) must be supplemented by
the non-relativistic limit of the Einstein-aether action (\ref{Saether}).
For simplicity, in the rest of this section we will
restrict to the case 
$c_2=c_3=c_4=0$ and describe the results to leading order in $c_1\ll 1$.
 This limit 
 shows already the key features of the generic case, to be considered
in Sec.~\ref{ssec:lin}. In this approximation, the action
(\ref{Saether}) at post-Newtonian order reads 
\be
\label{SaetherNew}
S_\text{\ae}=
\frac{M_0^2}{2}\int \di^4x\big[4\phi\D\psi-2\psi\D\psi
+c_1 u^i\Delta u^i
\big]\;.
\ee
Additionally,
we will assume $c_1\lesssim Y$. This is the most interesting regime
from the phenomenological perspective since we do not expect
to obtain bounds on the  LV in DM
matter sector that would be much stronger than the bounds on
$c_1$. 

The equations of motion following from
(\ref{fluidNew}) and (\ref{SaetherNew}) are\footnote{In this and only
  in this section
  dot denotes the total (material) derivative with respect to $t$.}  
\bseq
\label{Neweqs}
\begin{align}
\label{1peqNew}
&(1-Y)\,\dot{V}^i+\pd_i\phi+
Y(\pd_t u^i+V^j\pd_j u^i-V^j\pd_i u^j+u^j\pd_i u^j)=0\;,\\
&\psi=\phi\;,\\
\label{Neweq}
&2M_0^2\Delta\phi=\rho\;,\\
\label{aethNeweq}
&M_0^2c_1\Delta u^i=Y\rho\;(u^i-V^i)\;.
\end{align}
This system is closed by adding the continuity equation for DM, 
\be
\label{continuity}
\pd_t\rho+\pd_i(\rho V^i)=0\;.
\ee
\eseq

Consider a spherical DM halo of size $R_h$, constant density
$\rho_h$ and moving as a whole with velocity ${\bf V}_h$ with respect to
the preferred frame. According to Eq.~(\ref{aethNeweq}) the aether
perturbations inside the halo acquire the effective mass
\be
\label{meffh}
m_\mathrm{eff}^2=\frac{Y\rho}{M_0^2 c_1}\;.
\ee
Let us first assume that the halo is small,
\be
\label{Rhsmall}
R_h\ll m_\mathrm{eff}^{-1}\;,
\ee
so that the range of the aether interactions exceeds the halo
size. Note that this condition is equivalent to 
\be
\label{weakphi}
\phi_{h,s}\ll c_1/Y\;,
\ee
where $\phi_{h,s}\sim\rho_h R_h^2/M_0^2$ is the gravitational
potential at the surface of the halo. Then in (\ref{aethNeweq}) one can neglect the
$u^i$-term on the r.h.s. and obtain for the aether field produced by
the halo,
\be
\label{uweak}
u^i=-\frac{2Y V_h^i}{c_1}\,\phi_h({\bf x},t)\;,
\ee
where we have expressed the result in terms of the halo's
gravitational potential
\be
\label{phih}
\phi_h({\bf x},t)=-\frac{\rho_h R_h^3}{6M_0^2|{\bf x}-{\bf V}_ht|}\;.
\ee
We now study the motion of a test DM particle in the field of
the halo. Substituting (\ref{uweak}) into (\ref{1peqNew}) we obtain
for the particle's acceleration 
\be
\label{v2dot}
\dot{V}_{tp}^i=-\frac{\pd_i\phi_{h}}{1-Y}
+\frac{2Y^2}{(1-Y)c_1}
\big[-V_{tp}^jV_{h}^j\pd_i\phi_{h}
+V_{h}^i V_{tp}^j\pd_j\phi_{h}-
V_{h}^i V_{h}^j\pd_j\phi_{h}\big]\;.
\ee 
One observes two modifications compared to the standard case. First,
the gravitational acceleration towards the halo is enhanced by the
factor $1/(1-Y)$ due to the reduction of the inertial mass of the test
particle. This effect will play a key role in what
follows. Second, the combination in the square brackets gives rise to
the velocity-dependent interaction\footnote{Note that this interaction
  violates the (naive) third Newton's law: e.g. for a two-body system 
$m_{(1)}\dot{V}_{(1)}^i+m_{(2)}\dot{V}_{(2)}^i\neq 0$. This is
consistent with the observation \cite{Blas:2010hb} 
that in the presence of velocity dependent
forces the conserved momentum contains additional 
contributions, 
depending both on the velocities and the distance between the bodies 
(see also \cite{Will:Book}).} 
discussed in \cite{Blas:2010hb}.
The latter interaction is important as compared 
to the change in the gravitational acceleration
only if the halo and the particle
move fast enough,
\be
\label{vlarge}
|{\bf V}_{h,tp}|\gtrsim \sqrt\frac{c_1}{Y}\;.
\ee 
We are going to see (cf. (\ref{eq:vestimate}))  that the bulk
velocities of the linearized 
cosmological perturbations never satisfy the condition (\ref{vlarge}), 
so that the
velocity-dependent forces do not play a significant role in the
evolution of the Large Scale Structure. However, they may influence
the dynamics of non-linear structures such as galaxies and galaxy
clusters. We leave the study of this issue for the future.

Now consider the case when the size of the halo is larger than the inverse of the
aether mass in its interior,
\be
\label{Rhlarge}
R_h\gg m_\mathrm{eff}^{-1}\;.
\ee
Equivalently, 
\be
\label{strongphi}
\phi_{h,s}\gg c_1/Y\;.
\ee
Due to the Yukawa screening, the aether field is frozen at the minimum
of its potential, $u^i=V_h^i$ over most of the halo volume, deviating
from this value only in a surface layer of width
$m_\mathrm{eff}^{-1}$. Neglecting this deviation, to leading order
 the aether field outside the halo is
\be
u^i=\frac{R_h V_h^i}{|{\bf x}-{\bf V}_ht|}\;.
\ee 
Clearly, this is smaller than (\ref{uweak}) provided
(\ref{strongphi}) holds. Thus the velocity-dependent interaction
of a test particle with the halo  
is
suppressed in this case.
One notices the similarity with the chameleon mechanism
\cite{Khoury:2003aq} that screens scalar interaction in certain
scalar-tensor gravity models. However, it should be stressed that the
motion of a test particle is still modified compared to GR due to the
renormalization of the gravitational acceleration (cf. \eqref{v2dot}).

Finally,  instead of the test particle one can consider another large
halo, such that
its own gravitational potential satisfies (\ref{strongphi}). Let us
denote its velocity by ${\bf V}_{h1}$. Irrespectively of the aether
configuration outside, in the interior of this halo the aether will
coincide with $V^i_{h1}$. Substituting this into (\ref{1peqNew}) we
obtain  the standard equation,
\be
\label{standeq}
{\dot{V}}^i_{h1}=-\pd_i\phi_h\;.
\ee
One concludes that for large halos the deviations from GR are
completely screened\footnote{One may wonder if this screening
  mechanism can work for ordinary matter and help to relax some
  constraints on LV in the Standard Model. Unfortunately this is not
  the case as in the Standard Model LI is tested for individual
  elementary particles.}.

\subsection{Jeans instability}
\label{ssec:jeans}

In this subsection we analyze how the above peculiarities of DM 
dynamics in the presence of LV affect the development
 of the Jeans
instability. For simplicity, we will consider the universe filled
exclusively with DM; the more realistic case including the
cosmological constant, baryons and radiation will be studied in 
subsequent sections. Following the standard treatment of the Jeans
instability in the Newtonian limit, we take the spherically symmetric
Ansatz for the background values of the fields which we
denote with an over-bar. In this case the solution to Eqs.~(\ref{Neweqs}) reads
\be
\label{Hubbleflow}
\bar V^i=\bar u^i=H(t)\,x^i~,~~~~~~
\bar\phi=\frac{\bar\rho(t)}{12M_0^2}|{\bf x}|^2\;,
\ee 
where the Hubble parameter $H$ and the density $\bar\rho$ are related by  
\bseq
\label{cosmeqs}
\begin{align}
\label{Fried}
&\dot H+H^2=-\frac{\bar\rho}{6M_0^2}\;,\\
&\dot{\bar\rho}+3H\bar\rho=0\;.
\end{align}
\eseq
These are the standard equations of GR. In particular,
introducing the scale factor
\be
\label{scalea}
a(t)=\exp\left[\int^t H(t')\di t'\right]
\ee
and integrating Eqs.~(\ref{cosmeqs}) once, we obtain the Friedmann
equation 
\be
\label{eq:FriedII}
H^2=\frac{\bar \rho}{3M_0^2}+\frac{\kappa}{a^2},
\ee
with $\bar\rho\propto a^{-3}$. Thus at the level of the homogeneous
cosmology the model reproduces\footnote{As will be
  discussed in Sec.~\ref{sec:fluid}, there is actually a mismatch
  between the gravitational constant
  appearing in the Friedmann equation and the locally measured value (cf.
  \eqref{Gcosm}). This mismatch is of order $c_a$ and is
  not captured by the approximation used in this section.} GR. 
In what follows we will set
the spatial curvature to zero, $\kappa=0$.

The next step is to write  down the equations for the linear
perturbations. We obtain
\bseq
\begin{align}
&(1-Y)(\pd_tV^i+Hx^j\pd_jV^i+HV^i)+\pd_i\phi
+Y(\pd_tu^i+Hx^j\pd_ju^i+Hu^i)=0\;,
\label{linNew1}
\\
\label{linNew2}
&2M_0^2\Delta\phi=\bar\rho\,\delta\;,\\
\label{linNew3}
&M_0^2c_1\Delta u^i=Y\bar\rho\, (-V^i+u^i)\;,\\
\label{linNew4}
&\pd_t\delta+Hx^i\pd_i\delta+\pd_iV^i=0\;,
\end{align}
\eseq
where for simplicity we denote  the perturbations by the same letters as the total quantities.
We have introduced the density contrast
$\delta=\delta\rho/\bar\rho$. 
Performing the Fourier decomposition
\be
\delta\mapsto\e^{i{\bf kx}/a(t)}\delta \;,~\text{etc.}\;,
\ee
restricting to the longitudinal part
\be
\label{eq:scalar_v}
V^i=ik_i v~,~~~~u^i=ik_i u\;,
\ee
and solving (\ref{linNew2}), (\ref{linNew4}) for $\phi$ and $v$ 
we are left with two equations
\bseq
\begin{align}
\label{FFour1}
&(1-Y)(\pd_t^2\delta+2H\pd_t\delta)
-\frac{3H^2}{2}\,\delta
+Y\frac{k^2}{a}(\pd_tu+Hu)=0\;,\\
\label{FFour2}
&\bigg[c_1M_0^2\frac{k^2}{a^2}+Y\bar\rho\bigg]\,u
=\frac{Y\bar\rho\, a}{k^2}\,\pd_t\delta\;.
\end{align}
\eseq
Here we have used the Friedmann equation (\ref{eq:FriedII})  to
express the factor in the 
second term of (\ref{FFour1}) through the Hubble parameter.

The form of Eq.~(\ref{FFour2}) suggests to distinguish two
regimes. For short waves, 
\be
\label{klarge}
\frac{k^2}{a^2}\gg \frac{Y\bar\rho}{M_0^2c_1}\;,
\ee
the aether perturbation reads
\be
u=\frac{Y\bar\rho\, a^3}{M_0^2 c_1 k^4}\pd_t\delta\;.
\ee 
Note that the inequality (\ref{klarge}) is nothing but the condition for the unscreened regime,
Eq.~(\ref{Rhsmall}), with the size of the halo replaced by the physical
wavelength $a/k$ of the perturbation. Upon substitution into
(\ref{FFour1}) the aether gives only a small correction that can be
neglected. This can be understood as neglecting the velocity-dependent
interaction. Indeed, one estimates the velocity perturbation as (see
(\ref{linNew4})), 
\be
\label{eq:vestimate}
|{\bf V}| \sim H\delta\frac{a}{k}\ll H\delta 
\sqrt{\frac{M_0^2c_1}{Y\bar\rho}}\sim 
\delta\sqrt{\frac{c_1}{Y}}\ll
\sqrt{\frac{c_1}{Y}},
\ee 
where in the third relation we used the background Friedmann
equation; this is below the limit (\ref{vlarge}). Thus for the density
contrast we obtain the equation,
\be
\pd_t^2\delta+2H\pd_t\delta
-\frac{3H^2}{2(1-Y)}\,\delta=0\;. 
\ee
Its growing solution has the form
\be
\label{eq:Ngrowing}
\delta\propto t^\gamma~,~~~~~
\gamma=\frac{1}{6}\bigg[-1+\sqrt{\frac{25-Y}{1-Y}}\bigg]\;,
\ee
Clearly, for $Y>0$ the mode grows faster than in GR where $\gamma=2/3$.

In the opposite regime of long modes,
\be
\label{ksmall}
\frac{k^2}{a^2}\ll \frac{Y\bar\rho}{M_0^2c_1}\;,
\ee
the aether takes the form
\be
u=\frac{a}{k^2}\;\pd_t\delta\;.
\ee
The corresponding contribution in (\ref{FFour1}) combines with the
first term and restores the standard equation for the density
perturbations in GR as expected in the screened regime,
\be
\pd_t^2\delta+2H\pd_t\delta
-\frac{3H^2}{2}\,\delta=0\;,
\ee 
whose solution is $\delta\propto t^{2/3}$.

The following picture emerges from the above results. For a mode with
given $k$ the effects of LV are screened as long as the inequality
(\ref{ksmall}) is satisfied and the mode behaves exactly as in GR. As
the universe expands, this inequality breaks down\footnote{Recall that
$\bar\rho$ decays as $a^{-3}$ so that the r.h.s. of (\ref{ksmall})
decreases faster than the l.h.s.} and the
mode enters into the unscreened regime, where its growth is
accelerated up to (\ref{eq:Ngrowing}). Physically, this enhancement is
due to the increase in the gravitational acceleration of the DM
particles in this regime, see Eq.~(\ref{v2dot}),
which in turn is due to the change in their inertial mass.
We will see in 
Sec.~\ref{sec:CosmoPert} that this picture coincides with the results
of the full relativistic analysis in the appropriate limit 
(subhorizon modes, matter
dominated epoch).

Before closing this section, we mention that a straightforward
analysis of the transverse (vector) modes of the perturbations shows
that these modes decay in the same way as in GR both in the long-
and short-wave regimes. We will not consider the vector modes in what
follows. 


\section{Lorentz violating dark matter: relativistic fluid}
\label{sec:fluid}

To systematically develop the consequences of the aether -- DM
interaction beyond the Newtonian limit it is convenient to use
a relativistic fluid description of DM. The most suitable
for our purposes is the pull-back formalism, which enables to describe
perfect fluids with arbitrary
equation of state using an effective action, 
see \cite{Andersson:2006nr} for a review and early references.
Since the reader may not be familiar with this formalism,
we give a
succinct review of it in the next subsection, closely following the
presentation of \cite{Dubovsky:2005xd}. We will specialize in 
the case of cold DM with vanishing pressure
and present  an effective action describing its interaction with 
the aether.  
In Sec.~\ref{ssec:eqs} 
we use this action to derive the equations of motion. 
Then in Secs.~\ref{ssec:hom},
\ref{ssec:lin} we analyze the consequences of the resulting model for
the background cosmology and derive the equations for 
linear perturbations around Friedmann--Lema\^itre--Robertson--Walker (FLRW) solutions.

\subsection{Effective action for Lorentz violating fluids}
\label{ssec:effact}

In the pull-back formalism, the fluid elements are labeled
by three scalar fields $\varphi^I(x)$, $I=1,2,3$. 
An essential property of a perfect fluid is the
invariance of its dynamics when its elements are moved around without
changing their volume. 
This  means that the description in terms of the 
fields  $\varphi^I(x)$ must be invariant under
volume-preserving reparameterizations
\be
\label{reparams}
\varphi^I\mapsto\tilde\varphi^I(\varphi^J)\;,~~~~~
\det\frac{\pd\tilde\varphi^I}{\pd\varphi^J}=1\;.
\ee
The scalar object with the minimal number of derivatives invariant under
(\ref{reparams}) that can be constructed from $\varphi^I$ and the
metric $g_{\m\n}$ is 
\be
\label{B}
B\equiv-\det{B^{IJ}}\;,
\ee
where
\be
\label{BIJ}
B^{IJ}=g^{\mu\nu}\pd_\mu\varphi^I\pd_\nu\varphi^J\;.
\ee 
The Lagrangian for the perfect fluid is thus an arbitrary function of this
quantity, 
\be
\label{Sfluid}
S_{fluid}=-\int \di^4x\sqrt{-g}\,f(B)\;.
\ee
To make contact with the standard quantities of
hydrodynamics let us consider the vector\footnote{The
  antisymmetric pseudotensors are 
  defined as $\epsilon^{0123}=-\epsilon_{0123}=-1$, $\epsilon_{123}=1$.}
\be
\label{vm}
v^\m\equiv-\frac{\epsilon^{\m\n\s\rho}}{6\sqrt{-gB}}\,
\pd_\n\varphi^I\pd_\s\varphi^J\pd_\rho\varphi^K
\epsilon_{IJK}\;.
\ee
This vector has unit norm, is future directed and has the property
that the fluid labels $\varphi^I$ remain constant along its integral curves
\be
v^\m \pd_\m \varphi^I=0, \ \quad I=1,2,3.
\ee
Thus it is identified with the fluid velocity.
It follows that independently of the equations of motion
\be
\nabla_\m(\sqrt{B}\,v^\m)=0\;,
\ee 
which implies that $\sqrt{B}$ should be interpreted as the conserved
number density $n$ of the fluid elements. 
Other useful relations are
\be
\label{eq:usef}
B_{IJ}\pd_\m\varphi^I\pd_\n\varphi^J=g_{\m\n}-v_\m v_\n\;,~~~~~~~
\pd_\n\varphi^I\pd_\rho\varphi^J\pd_\s\varphi^K\epsilon_{IJK}=
-\sqrt{-g\,B}\,v^\m\epsilon_{\m\n\rho\s}\;,
\ee
where $B_{IJ}$ is the inverse matrix of $B^{IJ}$.
The energy-momentum tensor (EMT) following from (\ref{Sfluid}) 
can be cast into the standard hydrodynamic form
\[
T_{\m\n}=(\rho+p)v_\m v_\n-p g_{\m\n}\;,
\]
where the density and pressure are given by 
\be
\label{Frho}
\rho=f\;, \quad p=2f' B-f\;.
\ee
It is conserved due to the equations of motion for the fields $\varphi^I$, 
\be
\label{eqsfluid}
\nabla_\m\big(f'(B)BB_{IJ}\nabla^\m\varphi^J\big)=0\;.
\ee
For
cold DM $p=0$ implying that $f$ takes the form,
\be
\label{fdm}
f_{dm}=m\sqrt{B}= m\, n\;,
\ee
where $m$ is the mass of the fluid elements.

It is clear how to generalize the action (\ref{Sfluid}) to include the
effects of LV. In the presence of the aether the action can also depend on
the invariant scalar product $u_\m v^\m$, 
\be
\label{Sfluidpr}
\tilde S_{fluid}=-\int \di^4 x \sqrt{-g}\, \tilde f(B,u_\m v^\m)\;.
\ee
Once the microscopic theory of DM is specified the 
function $\tilde f$ can in principle be
computed. In this work we adopt  a phenomenological approach
and consider a generic function satisfying reasonable conditions.
A natural assumption for DM is that the Lagrangian
is proportional to the particle number,
\be
\label{fLVdm}
\tilde f_{dm}(B,u_\m v^\m)=m\sqrt{B}\,F(u_\m v^\m)\;,
\ee 
where the function $F$ is subject to the normalization condition 
$F(1)=1$. It is easy to convince oneself 
that most cold DM scenarios are described by functions of the form
\eqref{fLVdm}. In particular,  this
assumption is  valid for DM composed of well-separated 
particles that weakly interact with each other.
 The function $F$ then
coincides with that appearing in the one-particle action considered in
Sec.~\ref{sec:Newton}.

\subsection{Equations of motion}
\label{ssec:eqs}

The equations of motion for the DM fluid are obtained by varying
(\ref{Sfluidpr}) with respect to $\varphi^I$ and read
\begin{equation}
\label{dmeq}
\nabla_\mu
\Biggl[\sqrt{B}B_{IJ}\nabla^{\mu}\varphi^{J}\bigl(F-(u_\lambda
v^{\lambda})F'\bigr)+F'\frac{\epsilon^{\mu\nu\lambda\rho}}{2\sqrt{-g}}u_{\nu}
\partial_{\lambda}\varphi^{J}\partial_{\rho}\varphi^{K}\epsilon_{IJK}\Biggr]=0\;.
\end{equation}
A more conventional form involving only
hydrodynamic variables emerges after contracting with
$\pd_\s\varphi^I$
and using Eqs.~\eqref{eq:usef},
\begin{align}
\label{dmeq1}
-\nabla_\mu\Big[\rho_{[dm]}\Big(\big(F-(u_\lambda v^\lambda)F'\big)v^\m v_\s
+F'v^\m u_\s\Big)\Big]+\rho_{[dm]}F'v^\m \nabla_\s u_\m=0\;,
\end{align}
where we introduced
\be
\label{rhodm}
\rho_{[dm]}\equiv m\sqrt{B}\;.
\ee

The equations of motion for the aether are slightly different depending on whether
one considers a generic aether or a khronon for the LV sector. In the former case, from the variation of the
combination of (\ref{Saether}) and (\ref{Sfluidpr}) one finds
\be
\label{aethereq}
\nabla_\m K^\m_{~\,\n}-c_4a^\rho\nabla_\n u_\rho-l u_\n
-\frac{\rho_{[dm]}}{M^2_0} F' v_\n=0\;,
\ee
where we made use of the following notations,
\be
\label{Kms}
K^\m_{~~\s}\equiv K^{\m\n}_{~~~\s\rho}\nabla_\n u^\rho\;,~~~~~~
a_\m\equiv u^\lambda\nabla_\lambda u_\m\,.
\ee
Contracting (\ref{aethereq}) with $u^\n$ and using the constraint
(\ref{unitnorm}) we obtain the expression for the Lagrange multiplier,
\be
\label{l}
l=u^\n\nabla_\m K^\m_{~\,\n}-c_4a^\rho a_\rho-\frac{\rho_{[dm]}}{M_0^2}
(u_\m v^\m)F'\;.
\ee 
When substituted back into (\ref{aethereq}), this yields,
\be
\label{aethereq1}
\mathcal P^{\rho\mu}
\bigg(\nabla_\nu K^\nu_{\phantom{\n}\mu}-
c_4\,a_\nu\nabla_\mu u^\nu 
-\frac{\rho_{[dm]}}{M_0^2}
 F' v_{\mu}\bigg)=0\;,
\ee 
where 
\be
\label{calP}
\mathcal P_\m^\n \equiv \delta_\m^\n-u_\m u^\n
\ee
is the projector on the directions orthogonal to $u_\m$. In the
khronon case the constraint is identically satisfied and thus the Lagrange multiplier is
absent. Varying with respect to $\s$, we find the equation of motion
which is essentially 
the divergence of (\ref{aethereq1}), 
\be
\label{kheq}
\nabla_\rho\bigg[\frac{\mathcal P^{\rho\mu}}{\sqrt{\nabla^\lambda\s\nabla_\lambda\s}}
\bigg(\nabla_\nu K^\nu_{\phantom{\n}\mu}
-\alpha\,a_\nu\nabla_\mu u^\nu 
-\frac{\rho_{[dm]}}{M_0^2}
 F' v_\m\biggr)\bigg]=0\;.
\ee

The EMT of DM is obtained by varying
(\ref{Sfluidpr}) with respect to the metric. In the case of
the Einstein-aether we obtain 
\be
\label{Taedm}
T_{[dm]\,\m\n}=\rho_{[dm]}\big(F-(u_\lambda v^\lambda)F'\big) v_\m v_\n\;.
\ee
In the khronometric model there is an extra contribution coming
from the variation of $u_\m$ when the metric is varied, 
\begin{equation}
\label{Tkhdm}
\tilde T_{[dm]\,\mu\nu}=
\rho_{[dm]}\Big[\big(F-(u_\lambda v^\lambda)F'\big) v_\m v_\n
+(u_\lambda v^\lambda)F'u_\m u_\n\Big]\;.
\end{equation}
The difference between the previous expressions disappears once we
consider the total EMT of 
the aether / khronon --- DM system. This reads in both
cases\footnote{The indices in round brackets are symmetrized as 
$
\ K_{(\m\n)}\equiv\frac{1}{2}(K_{\m\n}+K_{\n\m})\;,~~\text{etc.}
$},
\be
\label{Tdarktotal}
\begin{split}
T_{[\text{ae}]\,\m\n}+T_{[dm]\,\m\n}=
M_0^2\bigg[&2\nabla_\s K^\s_{~(\m}u_{\n)} 
-\nabla_\lambda(K_{(\m\n)}u^\lambda)-\nabla_\s( K^\s_{~(\m}u_{\n)})
+\nabla_\s (K_{(\m}^{~~\s}u_{\n)})\\
&-c_1\nabla_\m u^\lambda\nabla_\n u_\lambda
+c_1\nabla_\lambda u_\m \nabla^\lambda u_\n
-2c_4 u_{(\n}\nabla_{\m)}u^\s a_\s+c_4 a_\m a_\n\\
&+\frac{1}{2}g_{\m\n}K^\s_{~~\lambda}\nabla_\s u^\lambda
-u_\m u_\n u^\lambda\nabla_\s K^\s_{~~\lambda}
+c_4 a_\rho a^\rho u_\m u_\n\bigg]\\
&+\rho_{[dm]}\Big(\big(F-(u_\lambda v^\lambda)F'\big) v_\m v_\n
+(u_\lambda v^\lambda)F'u_\m u_\n\Big)\;,
\end{split}
\ee 
where in the aether case we have substituted the Lagrange multiplier
from (\ref{l}). The first three lines in (\ref{Tdarktotal}) correspond
to the pure 
aether EMT, while the last line represents the contribution
of DM and DM --- aether
interaction. Note that LV leads to the departure of the latter contribution
from the perfect fluid form. 

For the rest of the Universe we will consider the 
standard scenario with baryons,
radiation and a cosmological constant\footnote{As we want to 
isolate the effect of LV in the DM sector, we will not consider
other possibilities for dark energy.}. 
Those possess EMT's,
\be
\label{EMTother}
T_{[b]\,\m\n}=\rho_{[b]} v_{[b]\,\m} v_{[b]\,\n}~,~~~
T_{[\gamma]\,\m\n}=\rho_{[\gamma]}
\bigg(\frac{4}{3} v_{[\gamma]\,\m} v_{[\gamma]\,\n}
-\frac{g_{\m\n}}{3}\bigg)~,~~~
T_{[\Lambda]\,\m\n}=\rho_{[\Lambda]} g_{\m\n}\;,
\ee
where $\rho_{[s]}$, $v_{[s]\,\m}$ is the density and 4-velocity of the
corresponding component. The equations of motion for these components
are given by the covariant conservation of the EMT's 
(\ref{EMTother}). We neglect the energy
exchange between baryons and radiation, as well as the effects due to
 neutrinos.

\subsection{Background cosmology}
\label{ssec:hom}

The spatially homogeneous and isotropic  Ansatz reads
\be
\label{FRWgen}
\di s^2=a^2(\tau)(\di \tau^2-{\bf \di  x}^2)\;,~~~
u_0=v_{[s]\,0}=a(\tau)~,~~~u_i=v_{[s]\,i}=0~,~~~
\rho_{[s]}=\rho_{[s]}(\tau)\;,
\ee
where we have introduced the conformal time $\tau$. It is
straightforward to check that with this Ansatz Eq.~(\ref{aethereq1})
for the aether, or alternatively Eq.~(\ref{kheq}) for the
khronon\footnote{The khronon field itself can be taken as an arbitrary
monotonic function of time, e.g. $\s=\tau$.}, is
identically satisfied. Substituting (\ref{FRWgen}) into (\ref{dmeq1})
we find
\be
\label{dmdensity}
\rho_{[dm]}\propto a^{-3}\;,
\ee 
so the DM density behaves in the same way as in the
standard FLRW universe. Using
(\ref{Tdarktotal}) we obtain the Friedmann equation,
\be
\label{Friedmann}
\frac{{\dot a}^2}{a^4}
=\frac{8\pi G_{cosm}}{3}\big(\rho_{[dm]}+\rho_{[b]}
+\rho_{[\gamma]}+\rho_{[\Lambda]}\big)\;,
\ee
where dot denotes henceforth differentiation with respect to the
conformal time. This has the same form as in GR,  with the gravitational
constant renormalized due to the contribution of the aether (khronon)
EMT in the Einstein's equations (cf. \cite{Carroll:2004ai}),
\be
\label{Gcosm}
G_{cosm}=\frac{1}{8\pi M_0^2}\bigg[1+\frac{\beta+3\lambda}{2}\bigg]^{-1}\;.
\ee
Note that $G_{cosm}$ differs from the Newton constant (\ref{GNaether})
measured in local experiments. This difference is constrained by
the BBN considerations
\cite{Carroll:2004ai} 
\be
\label{BBNconstr}
|G_{cosm}/G_N-1|\lesssim 0.1\;,
\ee
which places a rather mild bound on the aether parameters.

\subsection{Linear perturbations around FLRW}
\label{ssec:lin}

We now turn to the analysis of the linearized cosmological
perturbations.  
We focus on  scalar perturbations\footnote{As we mentioned
in Sec.~\ref{ssec:jeans}, the vector perturbations decay as
usual 
inside the horizon, so we do not expect any significant effect due to them. 
For the analysis of 
effects at cosmological distances in the Einstein-aether model with
Lorentz invariant
DM see \cite{ArmendarizPicon:2010rs}. Tensor modes
behave as in GR with a modified speed of propagation \cite{Jacobson:2008aj,Blas:2011zd}.} and choose
to work in the conformal Newtonian gauge,
\be
\di s^2=a(\tau)^2\left[(1+2\phi)\di \tau^2
-\delta_{ij}(1-2\psi)\di x^i\di x^j\right].
\ee
Since the scalar sectors of the Einstein-aether and khronometric theories
are equivalent at the level of perturbations, it is enough
to study  the khronometric case with the parameters (\ref{khpar}).
 For the perturbations of the
aether vector, the velocities of various components and their
densities we write
\be
\label{pertbs}
u_i=a \,\pd_i \chi\;,~~~ v_{[s]\,i}=a\,\pd_i v_{[s]}~,~~~
u_0=v_{[s]\,0}=a\,(1+\phi)~,~~~
\rho_{[s]}=\bar\rho_{[s]}+\delta\rho_{[s]}\;,
\ee
where $[s]=\{[dm],[b],[\gamma]\}$ and the overbar denotes the background
values. Substituting these expansions into (\ref{Tdarktotal}) and
(\ref{EMTother}) we obtain the set of linearized Einstein's
equations\footnote{In these and all subsequent equations dot denotes
  {\em partial} derivative with respect to $\tau$.},
\bseq
\label{Einsts}
\begin{align}
\label{Einst00}
&2\Delta\psi -3\mathcal{H}(2+\alpha \mathcal B)\dot{\psi} -\alpha \Delta\phi+\alpha \Delta\dot{\chi}
+\alpha\mathcal{H}(1-\mathcal B)\Delta\chi\nonumber
\\ 
&\qquad\qquad\qquad\qquad~~~~~-\frac{a^2}{M_0^2}(\delta\rho_{[dm]}+\delta\rho_{[b]}+
\delta\rho_{[\gamma]}+2\phi[\bar\rho_{[dm]}+\bar\rho_{[b]}
+\bar\rho_{[\gamma]}+\rho_{[\Lambda]})=0\;,\\
\label{Einst0i}
&\dot{\psi}+\mathcal{H}\phi
+\frac{\alpha\,  \mathcal{C}^2\Delta\chi}{2+\alpha\mathcal B}
-\frac{a^2}{M_0^2(2+\alpha\mathcal B)}\bigg[\bar\rho_{[dm]}(Y\chi+(1-Y)v_{[dm]})+
\bar\rho_{[b]}v_{[b]}+\frac{4}{3}\bar\rho_{[\gamma]}v_{[\gamma]}\bigg]=0\;,\\
&\ddot{\psi}+\mathcal{H}
(\dot{\phi}+2\dot{\psi})+(2\dot{\mathcal{H}}+\mathcal{H}^2)\phi 
+\frac{2\Delta (\phi-\psi) 
+\alpha\mathcal B \Delta(\dot{\chi}+2\mathcal{H}\chi)}{3(2+\alpha \mathcal B)}-\frac{a^2\delta
  \rho_{[\gamma]}}{3(2+\alpha \mathcal B)M_0^2}=0\;,
\label{Einstijtrace}\\
\label{Einstijtracefree}
&\phi-\psi-\beta(\dot{\chi}+2\mathcal{H}\chi)=0\;,
\end{align}
\eseq
where we defined 
\be 
\label{defs}
\mathcal B \equiv\frac{\beta+3\lambda}{\alpha},\quad  
\mathcal{C}^2\equiv \frac{\beta+\lambda}{\alpha} \,, 
\quad \mathcal H\equiv \frac{\dot a}{a}\,,
\quad Y\equiv F'(1)\,.
\ee
We will assume in what follows that the
  parameters $\mathcal{B}$ and $\mathcal{C}$ are both of order one.
The linearized khronon and DM equations are
\bseq
\label{eqs_perturbch}
\begin{align}
&\ddot{\chi}+2\mathcal{H}\dot{\chi}-\mathcal{C}^2
\Delta\chi+\left[\dot{\mathcal{H}}(1-\mathcal B)
+\mathcal{H}^2(1+\mathcal B)+\frac{Y\bar\rho_{[dm]}a^2}{\alpha
  M_0^2}\right]\chi  
\notag\\
& \qquad\qquad\qquad\qquad~~~~~-\frac{Y\bar\rho_{[dm]}a^2}{\alpha M_0^2}v_{[dm]}-\dot{\phi}-\mathcal{H}(1+\mathcal B)\phi -\mathcal B\dot{\psi}=0\;,
\label{khronoeq}\\
&\label{darkmeq1}
\delta\dot\rho_{[dm]}+3\mathcal{H}\delta\rho_{[dm]}-
\bar\rho_{[dm]}(\Delta v_{[dm]}+3\dot\psi)=0\;,\\
\label{darkmeq}
&\dot{v}_{[dm]}+\mathcal{H}v_{[dm]}+\frac{Y}{1-Y}
(\dot{\chi}+\mathcal{H}\chi)-\frac{\phi}{1-Y}=0\;.
\end{align}
\eseq
From (\ref{khronoeq}) we see  that the parameter $\mathcal{C}$ has the
physical meaning of the 
velocity of the khronon waves.
In what follows we will often use the integrated form of the last equation,
\be
\label{eq:vgensol}
v_{[dm]}=-\frac{Y}{1-Y}\, \chi+\frac{1}{a(\tau)(1-Y)}
\int^\tau \di \tau'\, a(\tau')\,\phi(\tau',{\bf x})\;.
\ee
If the gravitational potentials are known, this can be used
  to eliminate $v_{[dm]}$ from (\ref{khronoeq}) and obtain an
  equation for $\chi$.
We observe that, as in the Newtonian limit, 
LV in DM is governed at the linear level by the single
 parameter $Y$. Following  the discussion in
Sec.~\ref{sec:Newton}, we will assume it to lie in the range $0\leq Y<1$. 

For baryons and radiation we have the standard hydrodynamic
equations,
\bseq
\label{matt}
\begin{align}
\label{matt1}
&\delta \dot{\rho}_{[b]}+3\mathcal{H}\delta\rho_{[b]}
-\bar{\rho}_{[b]} \big(\Delta
  v_{[b]}+3\dot{\psi}\big)=0\;, 
&\dot{v}_{[b]}+\mathcal{H}v_{[b]}-\phi=0\;,\\
\label{matt3}
&\delta \dot{\rho}_{[\gamma]}+4\mathcal{H}\delta\rho_{[\gamma]}
-\frac{4}{3}\bar{\rho}_{[\gamma]} \big(\Delta
  v_{[\gamma]}+3\dot{\psi}\big)=0\;, 
&\dot{v}_{[\gamma]}-\frac{\delta\rho_{[\gamma]}}{4\bar\rho_{[\gamma]}}-\phi=0\;.
\end{align}
\eseq
The equations presented above completely determine the
evolution of the linearized cosmological perturbations. Their exact solution can be
obtained only numerically. In the following section we find
the approximate analytic form of the solution in various dynamical
regimes. This will enable us to determine the qualitative influence of
LV on the power spectra. The numerical analysis is postponed
till Sec.~\ref{sec:numerics}.

\section{Cosmological perturbations: qualitative analysis}
\label{sec:CosmoPert}

\begin{figure}[htb]
\begin{center}
\begin{picture}(300,195)(20,25)
\put(0,0){\includegraphics[scale=0.3]{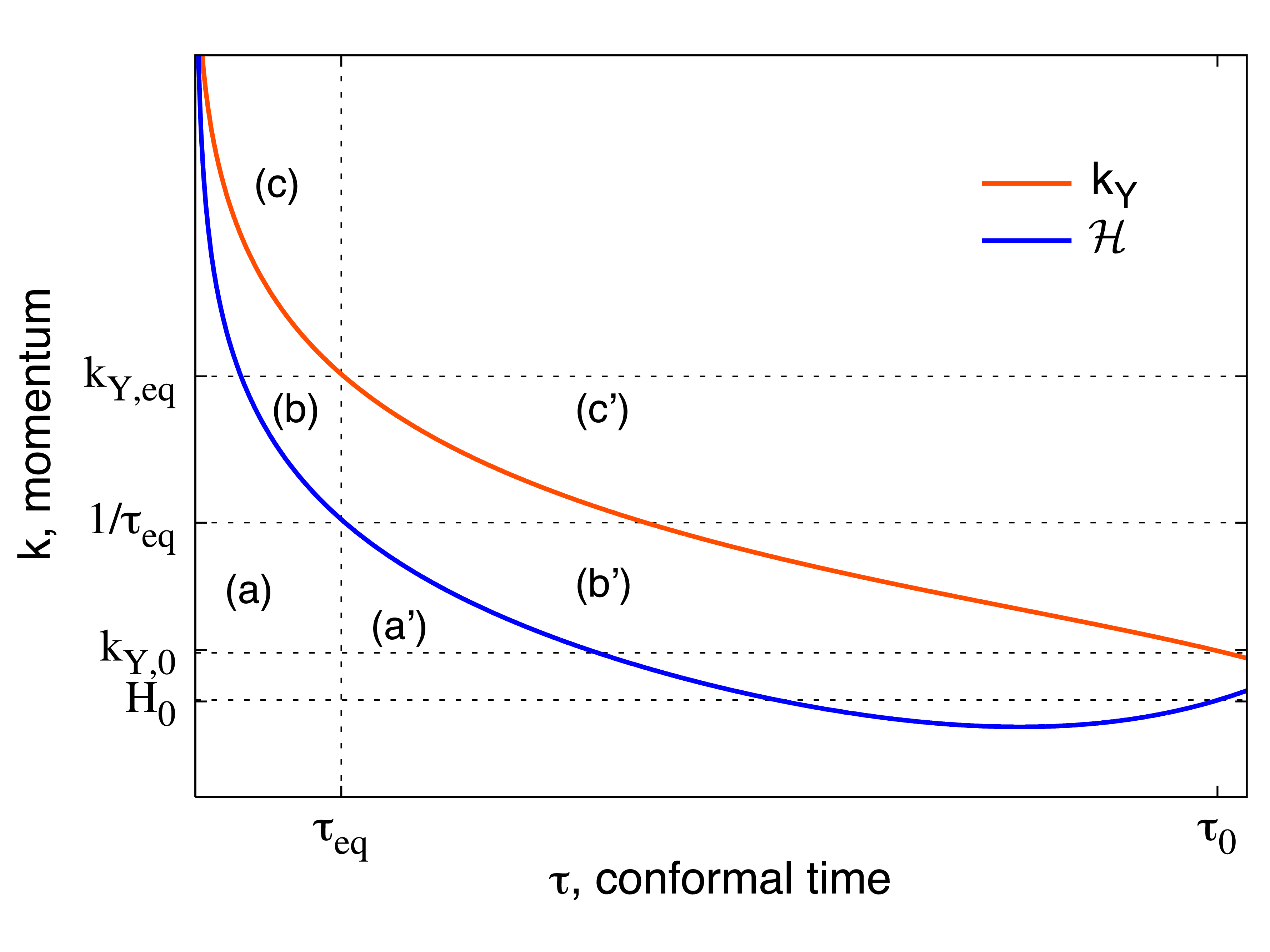}}
\end{picture}
\end{center}
\caption{Time dependence of the dynamical scales determining the
  evolution of the cosmological perturbations. $\tau_{eq}$ is the time
of radiation--matter equality and $\tau_0$ --- the present time. See
the main text for other notations.}
\label{fig:scales}
\end{figure}

Consider the mode with a given conformal wavenumber $k$. As usual, its
evolution will be different depending on whether $k$ is smaller or
larger than the expansion rate ${\cal H}$. Coupling between DM and
khronon gives rise to a second dynamical scale crucial for the mode
evolution. Substituting Eq.~(\ref{eq:vgensol}) into Eq.~(\ref{khronoeq}),
 we notice that the dynamics of the mode depends on the ratio between
 $k$ and
\begin{equation}
\label{ky}
k_{Y}\equiv\bigg(\frac{Y\bar\rho_{[dm]}a^2}{\alpha (1-Y)M^2_0}\bigg)^{1/2}\;.
\end{equation}
 This new scale is directly related to
the density of DM and determines
the critical wavenumber below which the effects of LV are screened
(see discussion in Sec.~\ref{sec:Newton}).
Factoring out the explicit time
dependence we can write,
\be
\label{ky1}
k_Y=\frac{k_{Y,0}}{\sqrt{a(\tau)}}~,~~~~
k_{Y,0}\equiv H_0\bigg[\frac{3Y\Omega_{dm}}{\alpha (1-Y)}
\bigg(1+\frac{\alpha\mathcal{B}}{2}\bigg)\bigg]^{1/2}\;,
\ee 
where $H_0$, $\Omega_{dm}$ are the present-day Hubble constant and 
DM density fraction; the current value of the scale
factor has been normalized to one, $a(\tau_0)=1$. Note that under the
assumption $\alpha\ll Y$ the present screening scale $k_{Y,0}$ is
parametrically higher than the Hubble rate $H_0$. The hierarchy
between $k_Y$ and ${\cal H}$ persists during the whole matter dominated
epoch and most of the radiation domination,
see Fig.~\ref{fig:scales}. 
Thus, the time evolution of a mode with given $k$ can be divided
into the following regimes:
\begin{itemize}
\item[{\it (a)}] superhorizon modes, $k<\mathcal{H}$, radiation dominated universe,
\item[{\it (a')}] superhorizon modes, $k<\mathcal{H}$, matter
  dominated universe,
\item[{\it (b)}] subhorizon modes with $\mathcal{H}<k<k_Y$, radiation
  dominated universe,
\item[{\it (b')}] subhorizon modes with $\mathcal{H}<k<k_Y$, matter
  dominated universe, 
\item[{\it (c)}] subhorizon modes with $\mathcal{H}<k_Y<k$, radiation
  dominated universe, 
\item[{\it (c')}] subhorizon modes with $\mathcal{H}<k_Y<k$, matter
  dominated universe. 
\end{itemize}

To find the solution of the linear equations we use the following
strategy. 
We first consider the formal limit
$\alpha,\beta,\lambda\to 0$ suggested by the constraints on the
khronon (aether)
parameters listed in 
Sec.~\ref{sec:aether}; at the same time we keep the parameter
$Y$ fixed. From (\ref{khronoeq}) one sees that in this
limit the khronon is tightly coupled to the DM velocity,
$\chi=v_{[dm]}$. Substituting this into the rest of  equations we
find that they take the standard GR form. Thus, to this approximation
all perturbations, in particular the gravitational potentials $\phi$
and $\psi$, behave as in GR. At the next step we allow for finite
$\alpha,\beta,\lambda$ (assuming all of them to be of the same order)  
and substitute the GR values for $\phi$, $\psi$
into Eqs.~(\ref{eqs_perturbch}) to find the khronon and
DM fluctuations. Finally, we find the corrections to $\phi$,
$\psi$ by inserting $\chi$, $v_{[dm]}$, $\delta\rho_{[dm]}$ into the
Einstein's equations (\ref{Einsts}).  

This logic has the caveat that in the limit $\alpha,\beta,\lambda\to 0$
the screening scale $k_Y$ diverges. Thus, the above perturbative scheme
is not applicable in cases 
{\it(c)} and {\it (c')} where the momentum $k$ is parametrically
large. 
Those cases will require a different approach.

\subsection{Regime {\it (a)}: $k<\mathcal{H}$ during radiation domination}
\label{ssec:5.1}

For the superhorizon modes, we can neglect all terms 
with spatial Laplacians and consider to leading order 
$\phi=\phi_{\gamma}=const$. At the radiation domination stage 
the scale factor depends linearly on the conformal time,  
\begin{equation}
\label{rdom}
a(\tau)=A_{\gamma}\tau, \quad A_{\gamma}=\sqrt{\Omega_{\gamma}}H_0\;,
\end{equation}
where $\Omega_\gamma$ is the present fraction of the radiation
density. Substituting this into Eqs.~(\ref{khronoeq}),
(\ref{eq:vgensol}) and setting the constant of integration in
(\ref{eq:vgensol}) to zero, which amounts to the absence of initial
velocities\footnote{Even if the initial velocities are
  present, their effect decays with time.}, we obtain,
\begin{align}
\label{supr}
&\ddot{\chi}+\frac{2\dot{\chi}}{\tau}
+\biggl[\frac{2\mathcal B}{\tau^2}+\frac{k^2_{Y,0}}{A_{\gamma}\tau}\biggr]\chi 
=\bigg[\frac{(1+\mathcal B)}{\tau}+\frac{k^2_{Y,0}}{2A_{\gamma}}\bigg]\phi_{\gamma}\;.
\end{align}
This equation has a solution
\be
\label{adiabr0}
\chi=\frac{\phi_\gamma}{2}\tau\;.
\ee
From (\ref{eq:vgensol}) we find that it corresponds to
\begin{equation}
\label{adiabr}
v_{[dm]}=\chi=\frac{\phi_{\gamma}}{2}\tau\;.
\end{equation}
This is the standard expression for the perturbation of 
DM velocity in the adiabatic mode and we see that the khronon evolves
coherently with the DM fluid. 
Note that this solution is completely insensitive to the
parameter $Y$ setting the strength of LV in DM.
The corresponding DM 
density contrast in the adiabatic mode is \cite{Ma:1995ey} (see also
Appendix~\ref{numerics}),
\be
\label{deltadmr}
\delta_{[dm]}\equiv\frac{\delta\rho_{[dm]}}{\bar\rho_{[dm]}}=
-\frac{3\phi_\gamma}{2}\;.
\ee
Inserting
(\ref{adiabr}) into the Einstein's equations (\ref{Einsts}) one can find the
corrections to the gravitational potentials induced by the
khronon. Clearly, these corrections are suppressed
by the parameters $\alpha,\beta,\lambda$ and we are not going to
analyze them in detail. Let us just note
that Eq.~(\ref{Einstijtracefree}) implies the appearance of a small anisotropic stress,
\be
\label{astress}
\frac{\phi-\psi}{\phi}=\frac{3\beta}{2}\;.
\ee

To determine whether the adiabatic mode (\ref{adiabr}) is an
attractor, we search for the solution of the homogeneous equation,
i.e. Eq. (\ref{supr}) with vanishing r.h.s.
This can be found in terms
of  Bessel functions,
\be
\label{freer}
\chi^{(a)}_{hom} \left( \tau \right) =\frac{\chi_{1}^{(a)}}{\sqrt\tau}\,
J_\nu\left(\frac{2k_{Y,0}}{\sqrt {A_\gamma}}\sqrt{\tau}\right)
+\frac{\chi_2^{(a)}}{\sqrt\tau}\,
Y_\nu\left(\frac{2k_{Y,0}}{\sqrt{A_\gamma}}\sqrt\tau\right)~,~~~~~
\nu=\sqrt{1-8{\cal B}}\;,
\ee
where $\chi^{(a)}_{1,2}$ are arbitrary coefficients. Recall that we
restrict to $0\leq Y<1$, in which case $k_{Y,0}$ is real and thus
$\chi^{(a)}_{hom}$ oscillates and decays at late times as $\tau^{-3/4}$. 
In this  case the adiabatic solution (\ref{adiabr0}) is an attractor. 
For negative $Y$, $k_{Y,0}$ is
imaginary and the solution (\ref{freer}) contains an
exponentially growing part signaling the instability of the
system. We have already encountered this instability when we studied
the Newtonian limit in Sec.~\ref{sec:Newton}. For the 
rest of the paper, we will only consider positive values of $Y$.

At early times, $\tau\ll A_\gamma/k_{Y,0}^2$, the first term in
(\ref{freer}) evolves as $\tau^q$ with
\be
\label{homr}
q=\frac{-1+\sqrt{1-8\mathcal B}}{2}\;.
\ee
For ${\cal B}<-1$ it grows faster than the adiabatic mode
(\ref{adiabr0}) and, depending on the initial conditions, may 
dominate over it. This corresponds to an intrinsic instability of the
khronon (not to confuse with the instability discussed in the previous
paragraph)
\cite{ArmendarizPicon:2010rs,Blas:2011en}. It is curious that the
coupling to DM cuts off this instability at $\tau\sim
A_\gamma/k^2_{Y,0}$, and makes the apparently growing mode decay. 
In this paper we will assume that the initial conditions for the
cosmological perturbations are sufficiently close to adiabatic, so
that this mode never becomes dominant.

\subsection{Regime {\it (a')}: $k<\mathcal{H}$ during matter domination}

During matter domination the scale factor grows as 
\be
\label{matdom}
a(\tau)=A_{m}\tau^2~,~~~~A_{m}=(\Omega_{dm}+\Omega_{b})H^2_0/4\;,
\ee
where $\Omega_b$ is the baryon density fraction today. 
At zeroth order, the superhorizon amplitude of
the gravitational potential is now $\phi_m=\frac{9}{10}\phi_\gamma$.
After substituting this into
Eqs.~(\ref{khronoeq}),~(\ref{eq:vgensol}) and neglecting the terms with
spatial Laplacians one obtains the following equation:
\begin{align}
\label{supm}
&\ddot\chi+\frac{4\dot\chi}{\tau}
+\bigg[2(1+3\mathcal B)+\frac{
    k_{Y,0}^2}{A_{m}}\bigg]\frac{\chi}{\tau^2} 
=\bigg[2(1+\mathcal B)+\frac{k^2_{Y,0}}{3A_{m}}\bigg]\frac{\phi_{m}}{\tau}\;.
\end{align}
Here we have
neglected the integration constant in (\ref{eq:vgensol}) which 
corresponds to a decaying contribution. The
adiabatic mode
\begin{equation}
\label{adiabmat}
v_{[dm]}=\chi=\frac{\phi_{m}}{3}\tau
\end{equation}
is a solution.
To check its stability consider the solution of the
homogeneous equation,
\be
\label{hommat}
\chi_{hom}^{(a')}(\tau)=\chi^{(a')}_1\tau^{r_+}+\chi^{(a')}_2\tau^{r_-}\;,~~~~~~~
r_\pm=\frac{1}{2}\bigg[-3\pm\sqrt{1-24\mathcal B-\frac{4k_{Y,0}^2}{A_{m}}}\bigg]\;.
\ee
For all parameters of interest the expression under the square root is
negative and $\chi_{hom}^{(a')}$ describes decaying oscillations. 
The adiabatic mode (\ref{adiabmat}) gives rise to the  
anisotropic stress
\be
\label{anidm}
\frac{\phi-\psi}{\phi}=\frac{5\beta}{3}\;.
\ee
Comparing to (\ref{astress}) and taking into account the ratio
between $\phi_m$ and $\phi_\gamma$ we see that the absolute difference
$(\phi-\psi)$ actually
stays constant at the transition from radiation to matter domination.

\subsection{Regimes {\it (b)}, {\it (c)}: 
$k>\mathcal{H}$ during radiation domination}
\label{ssec:bc}

The gravitational potentials decay rapidly inside the horizon during
the radiation domination, so we can neglect them in
Eqs.~(\ref{khronoeq}), (\ref{eq:vgensol}). The latter then takes the
form,
\be
\label{vsubr}
v_{[dm]}=-\frac{Y}{1-Y}\,\chi+\frac{v^{(0)}}{\tau (1-Y)}\;,
\ee
where $v^{(0)}$ is a constant. By matching
(\ref{vsubr}) approximately to (\ref{adiabr}) at horizon crossing $k\tau\sim 1$ we find 
$v^{(0)}\approx \phi_\gamma/2k^2$. Substitution into (\ref{khronoeq})
gives 
\be
\label{subr}
\ddot{\chi}+\frac{2\dot{\chi}}{\tau}
+\left[{\cal C}^2k^2+\frac{k^2_{Y,0}}{A_{\gamma}\tau}\right]\chi
=\frac{k_{Y,0}^2\phi_\gamma}{2A_\gamma k^2 \tau^2}\;,
\ee
where we have neglected the terms related to the Hubble parameter ${\cal H}$
in the square bracket. 

For $k\ll k_{Y,0}/\sqrt{A_\gamma\tau}$ (regime {\it (b)}) the solution reads,
\be
\chi(\tau) =\frac{\phi_\gamma}{2k^2\tau}+
\frac{\chi_1^{(b)}}{\sqrt\tau}
J_1\left(\frac{2k_{Y,0}}{\sqrt{A_\gamma}}\sqrt\tau
\right)+
\frac{\chi_2^{(b)}}{\sqrt{\tau}}
Y_1\left(\frac{2k_{Y,0}}{\sqrt{A_\gamma}}\sqrt{\tau}\right)\;.
\ee
This is a sum of a monotonically decreasing mode and damped
oscillations. The same structure is inherited by $v_{[dm]}$ through
Eq.~(\ref{vsubr}) and eventually, due to Eq.~(\ref{darkmeq1}),
leads to the standard logarithmic growth of the DM density
contrast (the LV effects are screened), accompanied by damped oscillations,
\begin{equation}
\label{delRD}
\delta_{[dm]}\approx-\frac{3\phi_\gamma}{2}-\frac{\phi_\gamma}{2}\log{(k\tau)}
+(\text{damped oscillations})\;.
\end{equation}

For $k\gg k_{Y,0}/\sqrt{A_\gamma\tau}$ (regime {\it (c)}) the formal 
perturbative expansion in
$\alpha,\beta,\lambda$ breaks down, as the corrections proportional to
$\alpha k^2$ in the Einstein's equations (\ref{Einsts}) may become large in this
regime (recall that
$k_{Y,0}$ is inversely proportional to
$\sqrt{\alpha}$). 
Nevertheless, we can rely on the fact that during the
radiation domination the DM perturbations are subdominant. 
It is reasonable to assume that this is also true for the khronon fluctuations
which are coupled to DM. Then the dynamics are still dominated by the
sound waves in the hot plasma that wash out the gravitational
potentials. Therefore, Eq.~(\ref{vsubr}) for the velocity of DM and
Eq.~(\ref{subr}) for the khronon still hold. Moreover, in the latter
equation one can neglect the second term in the square brackets and
the r.h.s. and obtain,  
\begin{equation}
\chi(\tau) =\chi_{1}^{(c)}\,\frac{\sin\left({\cal C}k\tau\right)}{\tau}
+\chi_{2}^{(c)}\,\frac{\cos\left({\cal C}k\tau\right)}{\tau}\;.
\end{equation}
Using (\ref{vsubr}), (\ref{darkmeq1}) we obtain for the density contrast,
\be
\label{delRD1}
\delta_{[dm]}=\const-\frac{\phi_\gamma}{2(1-Y)}\log{\tau}
+\frac{k^2 Y\chi_1^{(c)}}{1-Y}{\bf Si}({\cal C}k\tau)
+\frac{k^2 Y\chi_2^{(c)}}{1-Y}{\bf Ci}({\cal C}k\tau)
\ee
where ${\bf Si}(x)$ and ${\bf Ci}(x)$ are the integral sine and cosine, 
\[
{\bf Si}(x)= \int_0^{x}d\xi\;\frac{\sin\xi}{\xi},\quad {\bf
  Ci}(x)= \gamma + \ln(x)+ \int_0^xd\xi\;\frac{\cos\xi-1}{\xi}\;. 
\]
Notice that the coefficient in front of the logarithmic term has been
renormalized by the LV effects. The last two terms again describe damped oscillations.

The oscillations in the DM density and velocity field found above could
potentially provide
an interesting signature of the model. Unfortunately, the numerical
solution in Sec.~\ref{sec:numerics} shows that they are rather weak
and can leave a noticeable imprint only at very short wavelengths that
are presently in the non-linear regime. 
It is unclear whether such effects can be observed.

\subsection{Regimes {\it (b')}, {\it (c')}:  $k>\mathcal{H}$ 
during matter domination}

We finally consider the evolution of  subhorizon modes at the matter
dominated epoch. If one neglects the effects of LV, the
Newtonian potential stays constant, $\phi=\phi_{m,k}$, where the
subscript $k$ indicates the dependence on the wavenumber. Neglecting
in (\ref{eq:vgensol}) 
the integration constant that gives a rapidly decaying 
contribution and substituting the resulting expression for the
velocity into (\ref{khronoeq}) we obtain 
\be
\label{subm}
\ddot{\chi}+\frac{4\dot{\chi}}{\tau}+\left[{\cal C}^2k^2
+\frac{k_{Y,0}^2}{A_{m} \tau^2}\right]\chi
=\left[2(1+\mathcal B)+\frac{k_{Y,0}^2}{3 A_{m}}\right]\frac{\phi_{m,k}}{\tau}\;.
\ee
For $k\ll k_{Y,0}/\sqrt{A_m}\,\tau$ (regime {\it (b')}) the solution is
\be
\label{vchilong}
\chi=\left[\frac{6 A_{m}(1+\mathcal B)+k_{Y,0}^2}{12
    A_{m}+3k_{Y,0}^2}\right]
\phi_{m,k}\, \tau
+\chi_1^{(b')}\tau^{p_+}+\chi_2^{(b')}\tau^{p_-}~,~~~~
p_\pm=-\frac{3}{2}\pm\sqrt{\frac{9}{4}-\frac{k_{Y,0}^2}{A_m}}\;.
\ee
The definitions of $k_{Y,0}$, $A_m$ in (\ref{ky1}), (\ref{matdom})
imply that $k_{Y,0}^2/A_m\sim Y/\alpha$ is parametrically large. Thus
the last two terms in the above solution describe quickly decaying
oscillations and one is left with the attractor behavior 
(cf. (\ref{adiabmat})), 
\be
\label{vchilong1}
v_{[dm]}\approx \chi\approx\frac{\phi_{m,k}}{3}\tau\;.
\ee
The velocity perturbation has exactly the same form as in GR, so we
again conclude that in this regime the LV effects are
essentially screened. The corresponding density contrast is found from
(\ref{darkmeq1}) and exhibits the standard quadratic growth in time,
\be
\label{deltadmlong}
\delta_{[dm]}\approx -\frac{(k\tau)^2}{6}\phi_{m,k}\;.
\ee

It is instructive to go one step further and consider the corrections to
the gravitational potentials induced by the khronon\footnote{There are also
corrections to $\delta_{[dm]}$, $v_{[dm]}$ at the same order  which we do not consider here.}. From Eqs.~(\ref{Einstijtrace}), 
(\ref{Einstijtracefree}) we find
\be
\label{phipsicorr}
\phi=\phi_{m,k}\bigg(1+\frac{5}{84}(\beta+\lambda)(k\tau)^2\bigg)~,~~~~
\psi=\phi_{m,k}\bigg(1-\frac{5\beta}{3}
+\frac{5}{84}(\beta+\lambda)(k\tau)^2\bigg)\;.
\ee
These expressions are valid up to linear order in
$\alpha,\beta,\lambda$. We observe that the corrections, though small,
grow with time. Eventually,  when $k_Y$ red-shifts down to $k$, they
become of order $Y$: this follows from the estimate $(k_Y\tau)^2\sim
Y/\alpha$. The suppression of the corrections by the
small parameters $\alpha,\beta,\lambda$ disappears, which suggests the
break down of our perturbative scheme by the end of the regime {\it
  (b')}. Note however, that the anisotropic stress remains constant
and given by Eq.~(\ref{anidm})
during this stage.

For $k\gg k_{Y,0}/\sqrt{A_m}\,\tau$ (regime {\it (c')}) 
one can show that the corrections to the gravitational potentials are
not proportional to the expansion parameters $\alpha,\beta,\lambda$,
so the perturbative calculation fails.  
Thus we have to solve 
the coupled Einstein--khronon--DM equations self-consistently.
One notices that (\ref{Einstijtrace}),
(\ref{Einstijtracefree}), (\ref{khronoeq}), (\ref{darkmeq}) form a
closed system of equations for $\phi$, $\psi$, $\chi$,
$v_{[dm]}$. The educated guess of a power-law behavior yields the growing mode,
\bseq
\label{NPs}
\begin{align} 
\label{NPphi}
\psi&=\phi=\tilde\phi_{m,k}\; \tau^{\varkappa}\;,\\
\label{NPchi}
\chi&=\frac{k_{Y,0}^2\tilde\phi_{m,k}}{{\cal C}^2k^2A_m(\varkappa+3)}\,
\tau^{\varkappa-1}\;, \\
\label{NPv}
v_{[dm]}&=\frac{\tilde\phi_{m,k}}{(1-Y)(\varkappa+3)}\,\tau^{\varkappa+1}\;, 
\end{align}
\eseq
where 
\be
\label{NPpow}
\varkappa=-\frac{5}{2}+\sqrt{\frac{25}{4}+\frac{6Y}{1-Y}
    \frac{\Omega_{dm}}{\Omega_{dm}+\Omega_b}}\;, 
\ee
and we have neglected the corrections of order $\alpha,\beta,\lambda$. The mode
normalization $\tilde\phi_{m,k}$ can be determined by matching
(\ref{NPs}) to the solution before crossing the screening horizon
$k=k_Y$.

Substituting (\ref{NPv}) into (\ref{darkmeq1}) we find the behavior of
the DM density contrast,
\begin{equation}
\label{delMD2}
\delta_{[dm]}=-\frac{k^2\tilde\phi_{m,k}}{(1-Y)(\vk+3)(\vk+2)}\,
\tau^{\vk+2}\;. 
\end{equation}
One observes that $\delta_{[dm]}$ grows faster than in GR, implying the
accelerated growth of structure. The power-law (\ref{delMD2})
coincides with the results in Sec.~\ref{ssec:jeans} (see
Eq.~(\ref{eq:Ngrowing})) if we neglect the baryon contribution,
i.e. set $\Omega_b=0$, and take into account the relation between the
conformal and physical time, $\tau^3\propto t$.

It is interesting to study the perturbations of the baryonic
component. From (\ref{matt1}) we find,
\be
\label{delb2}
\delta_{[b]}\equiv\frac{\delta\rho_{[b]}}{\bar\rho_{[b]}}=
-\frac{k^2\tilde\phi_{m,k}}{(\vk+3)(\vk+2)}\,\tau^{\vk+2}\;.
\ee
Thus the baryon density contrast grows at the same rate as that of DM, 
but with an overall amplitude suppressed by a
$Y$-dependent factor,
\be
\label{bias}
\frac{\delta_{[b]}}{\delta_{[dm]}}=1-Y\;.
\ee
This behavior has a simple physical explanation: baryons do not
feel the enhanced gravitational acceleration experienced by DM. 
Therefore it takes them longer to react to the inhomogeneities
of the gravitational potential and their density perturbations stay
behind those of DM.

The anisotropic
stress is found from (\ref{Einstijtracefree}) and reads,
\be
\label{anis2}
\frac{\phi-\psi}{\phi}=\beta\cdot\frac{k_{Y,0}^2}
{{\cal C}^2 A_m k^2\tau^2}\;. 
\ee
Notice that the second factor on the r.h.s. is always smaller than 1
in the considered regime, so that the anisotropic stress is
still suppressed by $\beta$. Moreover, it decays with time as
$\tau^{-2}$.  

Finally, it is straightforward to check that the solution (\ref{NPs}),
(\ref{delMD2}), (\ref{delb2}) satisfies the remaining Einstein's
equations (\ref{Einst00}), (\ref{Einst0i}).

\subsection{Qualitative analysis of the power spectra}

The previous analysis allows us to sketch the qualitative picture of
the 
effects due to LV in DM in the observed 
power spectra of DM and baryon
density 
perturbations. From Fig.~\ref{fig:scales} we see that the modes with
different $k$ go through a different sequence of the dynamical regimes 
described before.  
\begin{figure}[tb]
\begin{center}
\begin{picture}(300,230)(20,25)
\put(0,0){\includegraphics[scale=0.3]{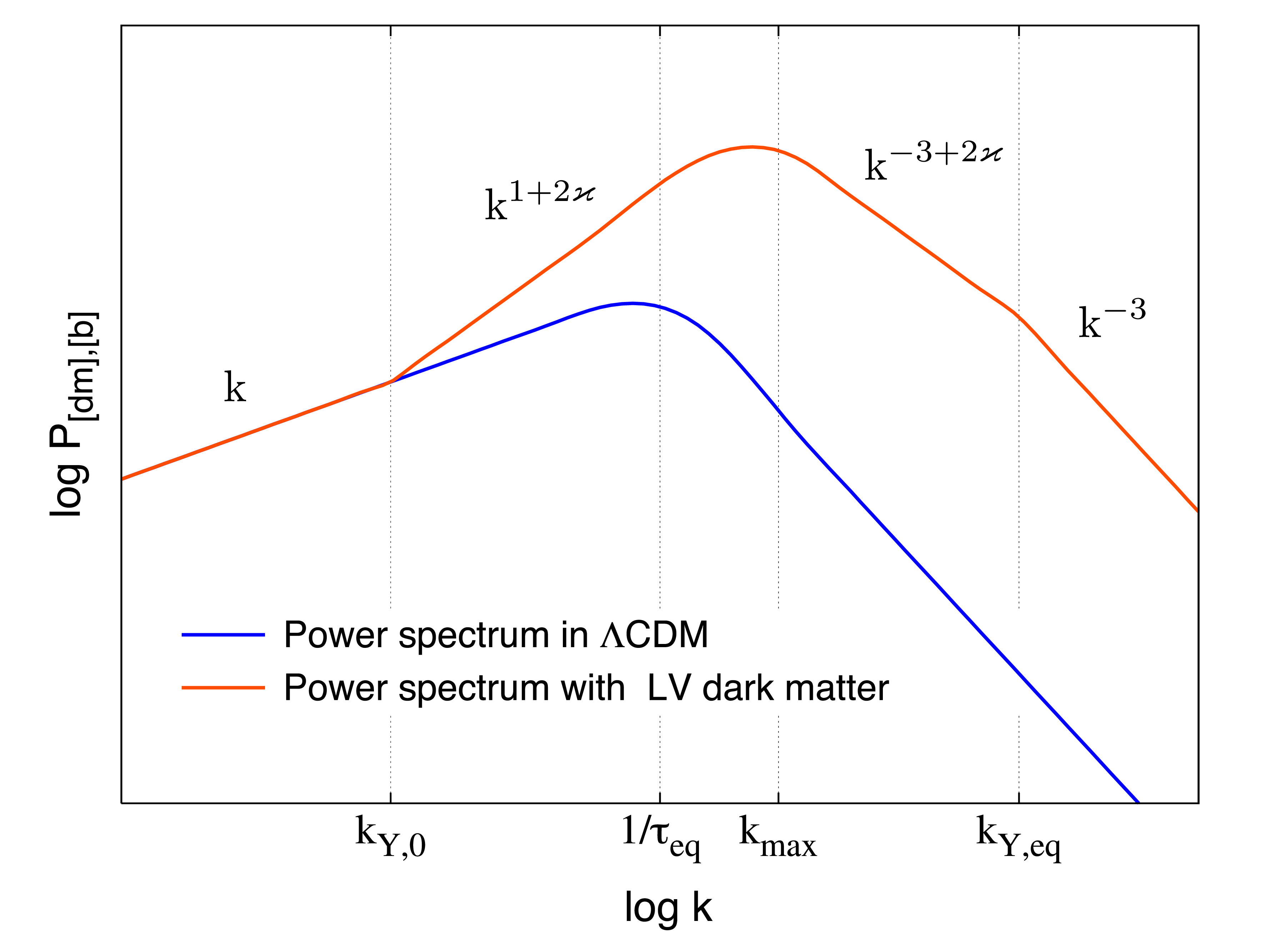}}
\end{picture}
\end{center}
\caption{Schematic representation of the matter power spectrum in the
  LV dark matter model (upper curve) compared to the power spectrum in
  $\Lambda$CDM (lower curve). The same qualitative behavior is common to the dark
  matter and baryons.
The quantities $k_{Y,0}$, $k_{Y,eq}$, $\vk$ are
  defined in (\ref{ky1}), (\ref{kyeq}), (\ref{NPpow}). The figure corresponds to the case
  $k_{Y,0}<1/\tau_{eq}$.}
\label{fig:pows}
\end{figure}
\begin{itemize}
\item 
The modes with $k<k_{Y,0}$ always stay in the regime where the LV
effects are essentially  screened. In this range the power spectrum has the
same dependence on $k$ as in $\Lambda$CDM.
\item
The modes with $k_{Y,0}<k<k_{Y,eq}$, where
\be
\label{kyeq}
k_{Y,eq}\equiv\frac{k_{Y,0}}{\sqrt{a(\tau_{eq})}}
=k_{Y,0}\sqrt{\frac{\Omega_{dm}+\Omega_b}{\Omega_\gamma}},
\ee
is the screening scale at the radiation--matter equality, enter into
the regime of unscreened LV at the matter dominated stage. Let us
denote the moment of the ``screening horizon'' crossing by $\tau_Y(k)$,
\[
\tau_Y(k)\equiv\frac{k_{Y,0}}{\sqrt{A_m}k}\;.
\]
Between $\tau_Y$ and $\tau_0$ the density contrasts exhibit the anomalous
growth (\ref{delMD2}), (\ref{delb2}). Thus they are enhanced by a
factor $(\tau_0/\tau_Y)^\vk$ leading to the increase of the power
spectra by a factor $\propto k^{2\vk}$ in this range of momenta.
\item 
Finally, the modes with $k>k_{Y,eq}$ enter in the unscreened regime
already during  radiation domination. Modulo the small damped
oscillations mentioned in Sec.~\ref{ssec:bc}, 
the net effect of LV on these modes is the overall enhancement by the
factor $(\tau_0/\tau_{eq})^{\vk}$. This factor does not depend on the mode
momentum, so the slope of the spectrum remains as in $\Lambda$CDM.
\end{itemize}
This picture is summarized in Fig.~\ref{fig:pows}. 
For the sake of the schematic representation we have assumed a
flat spectrum for the initial perturbations,
neglected the logarithmic growth of the perturbations during the
radiation domination together with the effects of the cosmological
constant. Note that the change in the slope of the power spectrum
depends only on the parameter $Y$ describing LV in DM,
while the range of scales where this change occurs is determined both
by $Y$ and the khronon parameters $\alpha$, $\beta$, $\lambda$.

\begin{figure}[htb]
\centering
\includegraphics[scale=0.8]{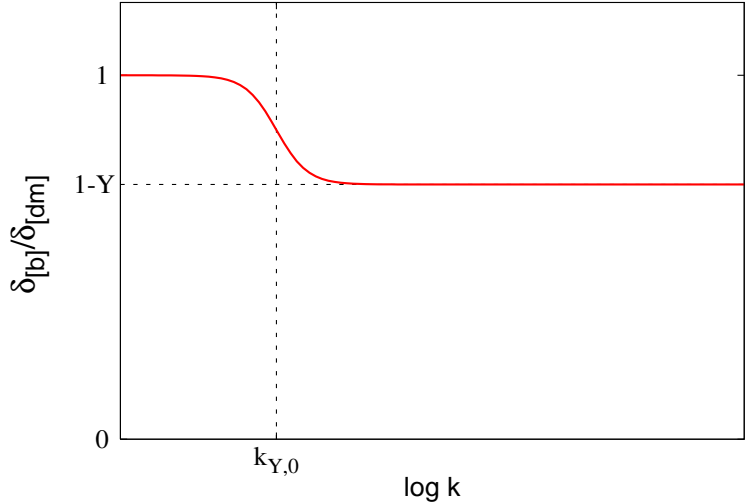}
\caption{Ratio between the density contrasts of baryons and dark
matter (qualitative plot).}
\label{fig:bias}
\end{figure}

Figure~\ref{fig:pows} corresponds to the situation when 
the ratio $Y/\alpha$ is moderately large, so that 
$k_{Y,0}$ is
smaller than $1/\tau_{eq}$, the Hubble rate at the radiation--matter
equality that sets the position of the maximum of the power spectrum
in $\Lambda$CDM. In this case the change in the slope of the power
spectrum leads also to the shift of its maximum. Indeed, in the
log-log scale the standard $\Lambda$CDM power spectrum in the vicinity
of the maximum can be written as
\[
\log P^{\Lambda CDM}=A_1-A_2(\log{k\tau_{eq}})^2\;,
\]
where $A_{1,2}$ are some constants. The additional enhancement due to
LV adds a linear contribution on top of it,
\[
\log P=\log P^{\Lambda CDM}+2\vk \log k\;.
\] 
Differentiating this expression with respect to $k$ we obtain the new
position of the maximum,
\be
\label{kmax}
\log{k_{max}}=\log{1/\tau_{eq}}+\vk/A_2\;.
\ee
We conclude that the shift is linearly proportional to $\vk$.

If $k_{Y,0}>1/\tau_{eq}$, which happens for an extreme hierarchy
between $Y$ and $\alpha$, 
the position of the maximum of the power spectrum is not modified.
However, the change in the slope is still
 present providing the signature of LV. 
 
The second qualitative
effect of LV in DM is the appearance of
the baryon --- DM bias (\ref{bias}) for the modes
undergoing the anomalous growth, i.e. all modes with wavenumbers
larger than $k_{Y,0}$. For smaller $k$ the density contrasts of the baryons
and DM are equal. 
This results in the scale dependence of the bias
shown in Fig.~\ref{fig:bias}.
It would be interesting to understand how this effect could be
constrained by
observations.

Finally, at non-zero value of the parameter $\beta$ the 
model is also characterized by the
anisotropic stress.
Its present-day value is independent of $k$ for 
$k<k_{Y,0}$ where it is given by Eq.~(\ref{anidm}).
At larger 
wavenumbers it falls off as $k^{-2}$ according to Eq.~(\ref{anis2}).

The qualitative analysis of this section is confirmed by the numerical
calculations, to which we proceed now.

\section{Cosmological perturbations: numerics}
\label{sec:numerics}

The study of the effects of LV in the DM sector yielded the modified
linearized equations for cosmological perturbations (\ref{Einsts}), 
(\ref{eqs_perturbch}), (\ref{matt}).
In this section we will solve these equations numerically for
different sets of LV parameters  
and confirm the modifications
in the growth of perturbations uncovered 
in Sec.~\ref{sec:CosmoPert}.
The density fractions of the
cosmological constant, DM, baryons and radiation are taken to
be  
\[
\Omega_{\Lambda}=0.75,\quad \Omega_{dm}=0.2,\quad \Omega_b=0.05,\quad
\Omega_{\gamma}=5\cdot 10^{-5}.
\]
We neglect the interaction between  baryons and  radiation as
well as the effects of neutrinos. As our goal is to compare the evolution
of perturbations in models of LV DM with $\Lambda$CDM,
we also find the evolution of perturbations in $\Lambda$CDM within the same
approximations\footnote{The corresponding equations follow from (\ref{Einsts}), (\ref{darkmeq1}), (\ref{darkmeq}) by
  setting $Y=\alpha=\beta=\lambda=0$, $\chi=v_{[dm]}$.}. The details of
the numerical procedure are presented in Appendix~\ref{numerics}. 

\begin{figure}[!htb]
\begin{center}
\includegraphics[width=0.45\textwidth]{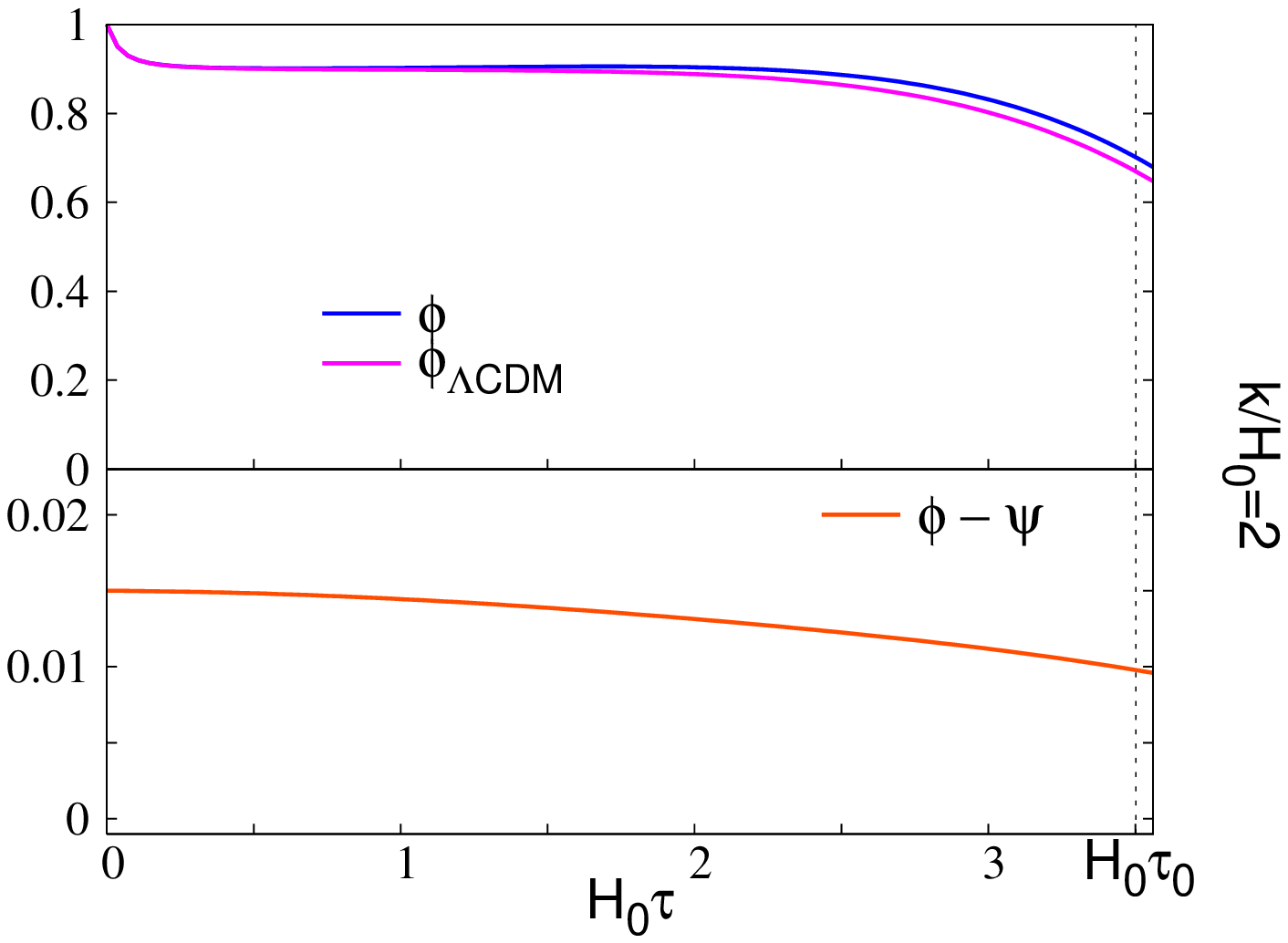}\qquad\quad
\includegraphics[width=0.45\textwidth]{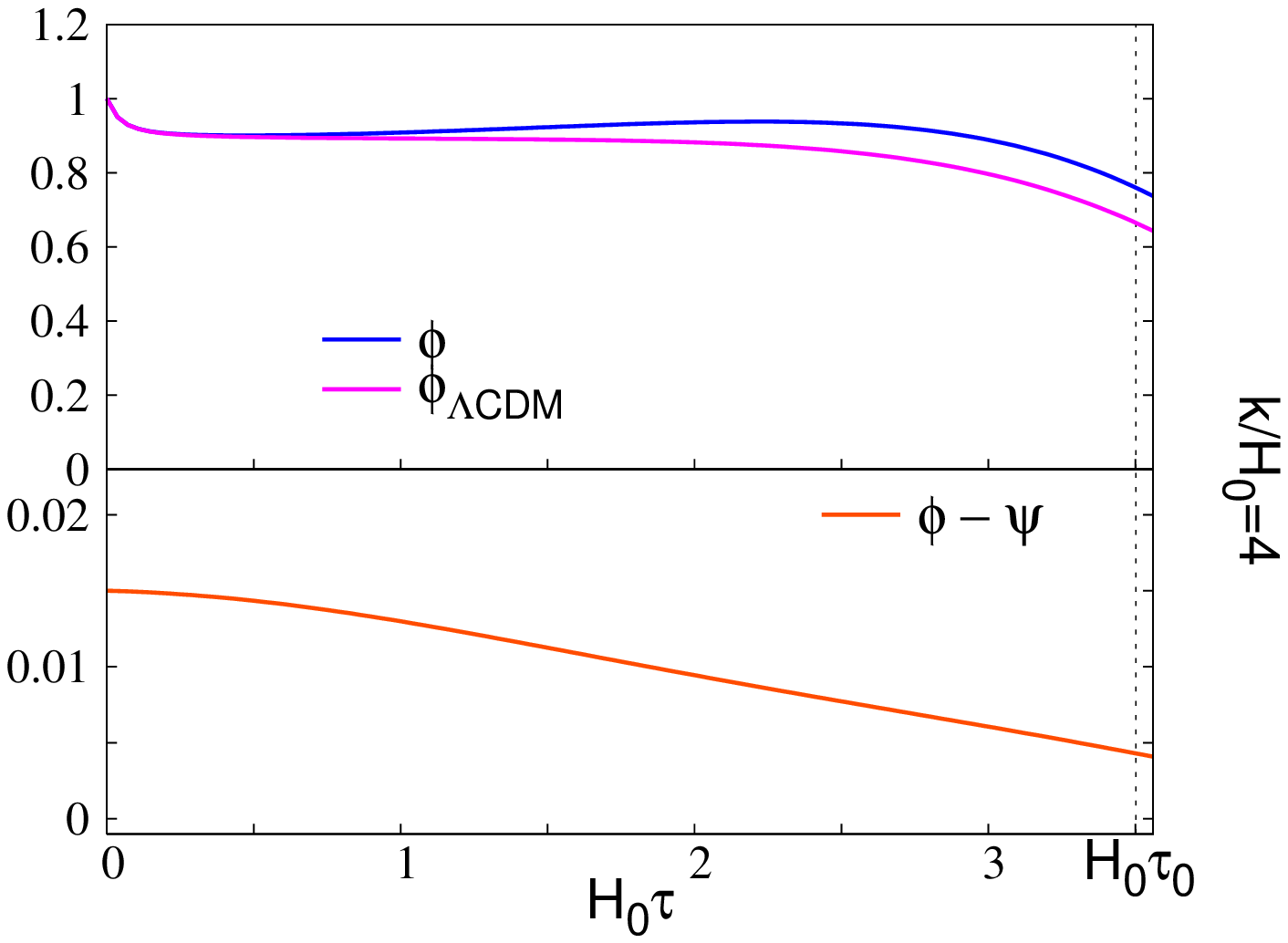}\\
\includegraphics[width=0.45\textwidth]{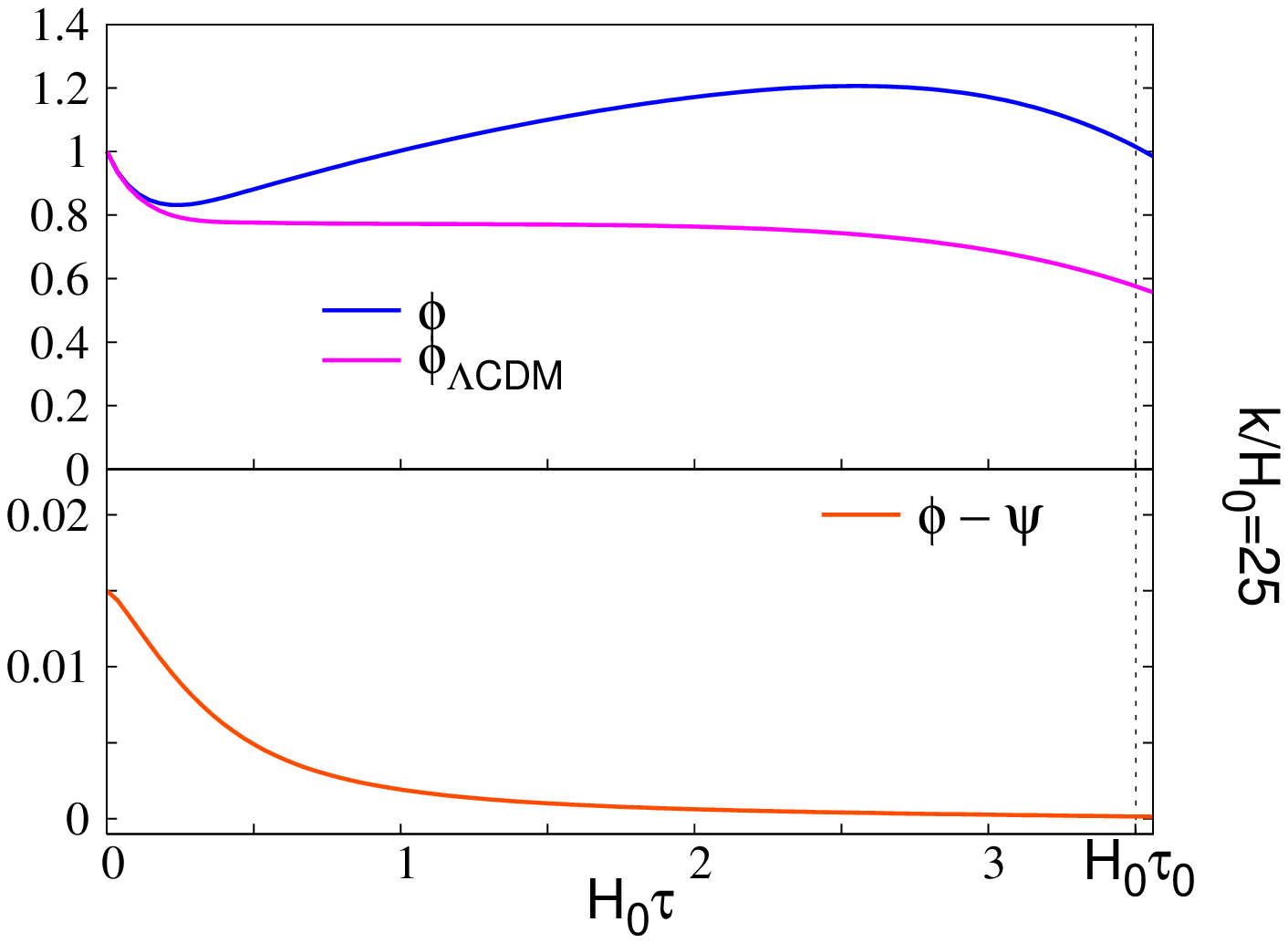}\qquad\quad
\includegraphics[width=0.45\textwidth]{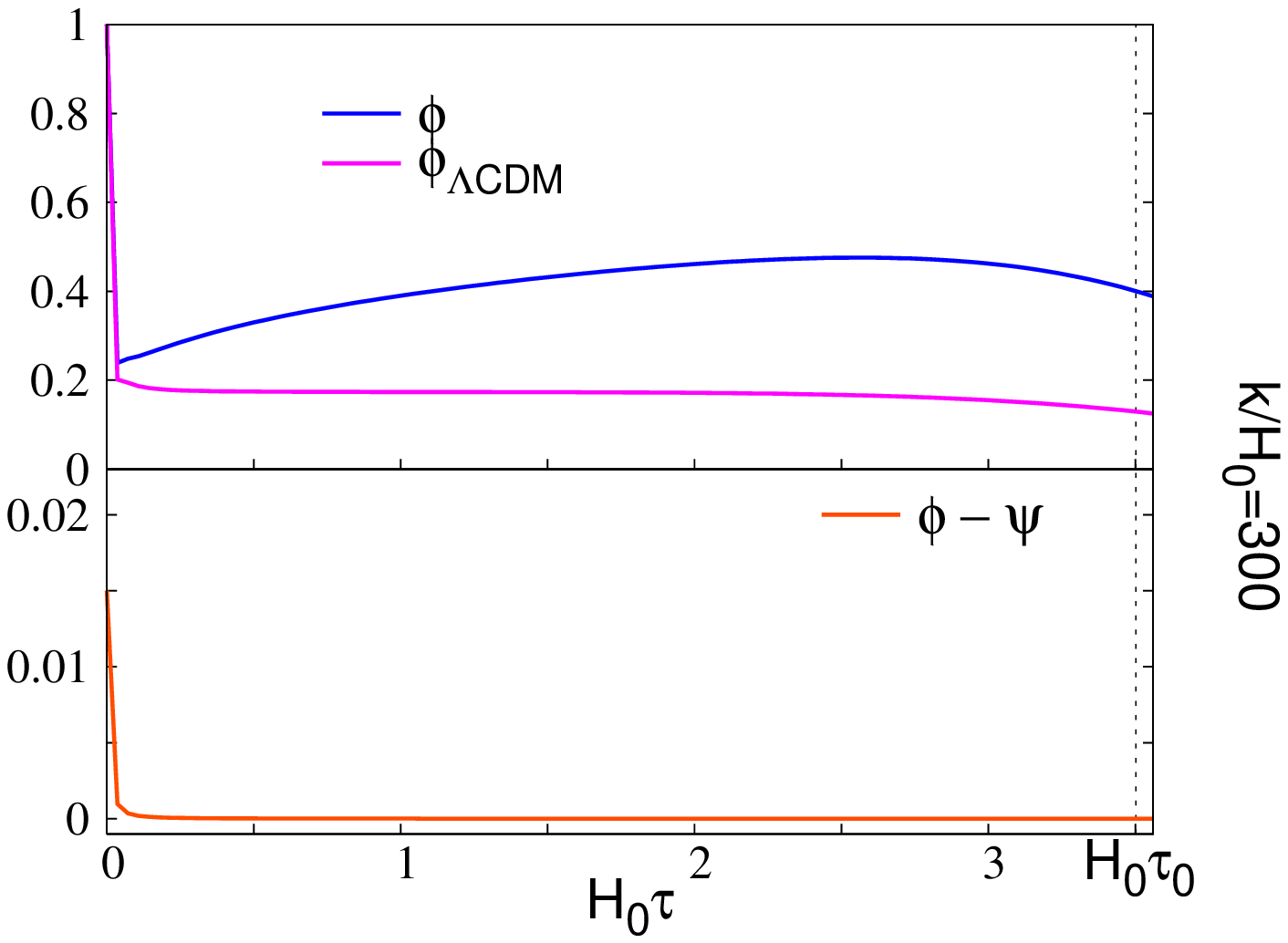}
\caption{Dependence of the gravitational potentials on conformal time for modes
  with momenta $k/H_0=2,\,4,\,25,\,300$. The LV parameters of the model are $\alpha=0.02$, 
$\beta=0.01$, 
$\lambda=0.01$, $Y=0.2$. The standard $\Lambda$CDM result is shown
for comparison. Dashed vertical lines mark the present conformal time
$\tau_0=3.5H_0^{-1}$. 
\label{Fig:2}
}
\end{center}
\end{figure}
The first set of parameters we consider is
\be
\label{choice1}
\alpha=0.02~,~~~\beta=0.01~,~~~\lambda=0.01~,~~~ Y=0.2\;.
\ee
This choice satisfies the PPN\footnote{Recall that $\alpha=2\beta$ case
  avoids the PPN bounds.}, BBN and gravitational radiation
constraints \cite{Blas:2010hb,Blas:2011zd}.
The initial conditions correspond to the adiabatic mode, and for
illustration purposes the 
initial value of $\phi$ is normalized to 1 for all modes. 
Figure \ref{Fig:2} shows the dependence of the Newton potential on
conformal time for the choice (\ref{choice1}) and for $\Lambda$CDM for
several modes with different values of momentum. The difference
$(\phi-\psi)$ is also shown. We see that perturbations of the gravitational potential are enhanced
at late times as compared to $\Lambda$CDM. The enhancement is
stronger for shorter modes that enter earlier into the regime where
the LV effects are not screened. Note that the present-day screening
scale corresponding to the choice (\ref{choice1}) is $k_{Y,0}=2.77\,H_0$.
On the other 
hand, the difference between the two gravitational potentials 
$(\phi-\psi)$ which is initially of order $10^{-2}$ decreases
once the mode enters inside the screening horizon. Note that the overall
amplitude $10^{-2}$ for long wavelength modes 
agrees with the estimate $(\phi-\psi)/\phi\sim\beta$ obtained in the
previous section.

\begin{figure}[htb]
\begin{center}
\includegraphics[width=0.6\textwidth]{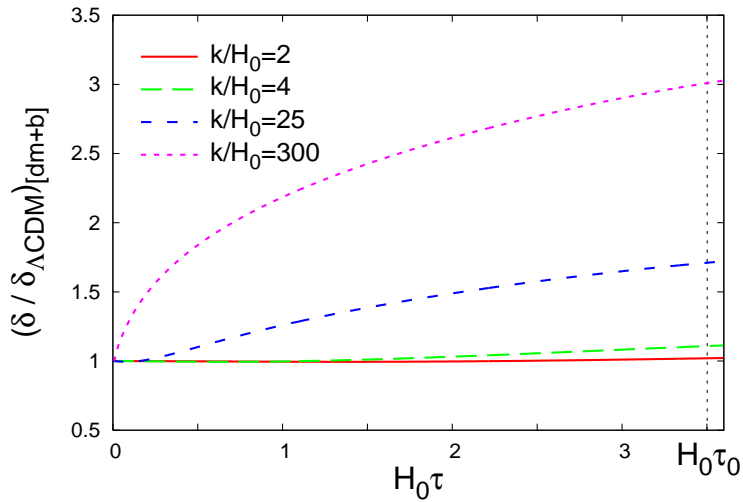}
\caption{Ratio of total density contrasts between the model 
with LV DM and 
$\Lambda$CDM versus time for several values of the
  mode momentum $k$. The parameters are the same
  as for Fig.~\ref{Fig:2}. $\tau_0$ is the present conformal time.  
\label{Fig:4}
}
\end{center}
\end{figure} 
In 
Fig.~\ref{Fig:4} we plot the total (DM plus
baryons) density contrast in the present model divided by its value
in $\Lambda$CDM. As
expected from the analysis of Sec.~\ref{sec:CosmoPert} (or simply
from the enhancement of the gravitational perturbations), the growth of
structures increases at recent times in a wide range of momenta. 
The relative effect is stronger for shorter modes and
can be large for our choice of parameters.

\begin{figure}[!htb]
\begin{center}
\includegraphics[width=0.45\textwidth]{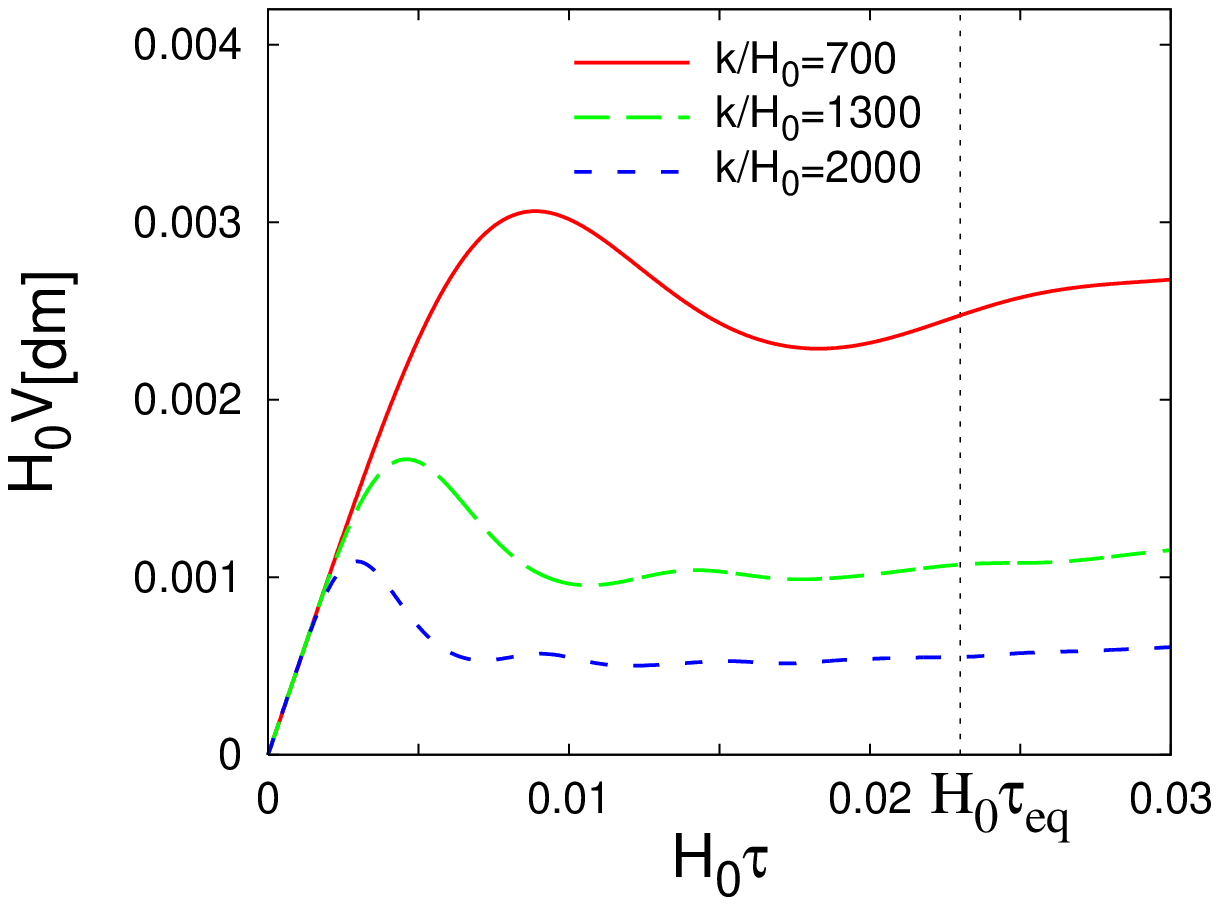}\qquad\quad
\includegraphics[width=0.45\textwidth]{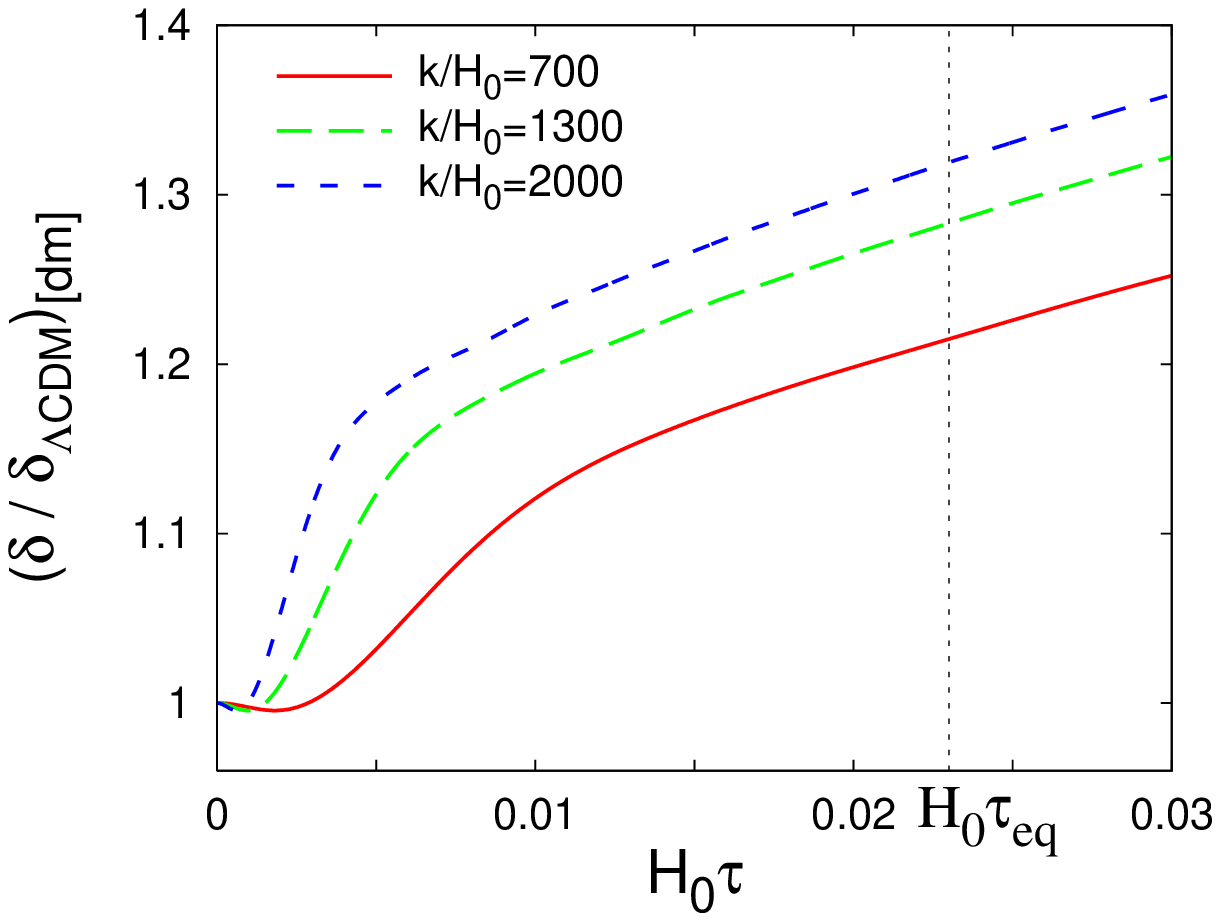}
\caption{Time-dependence of the longitudinal component of the DM 
velocity  (left
  panel) and the DM 
density contrast (right panel) for several subhorizon
  modes at the radiation-dominated epoch. The dashed vertical line
  marks the time of radiation -- matter equality, $\tau_{eq}\approx
  0.023 H_0^{-1}$. The parameters of the model are the same as for
  Fig.~\ref{Fig:2}.  
\label{Fig:oscillations}
}
\end{center}
\end{figure}
In Sec.~\ref{ssec:bc} we found that the 
subhorizon DM perturbations
exhibit damped oscillations during the radiation-dominated epoch. To
study this effect we plot in Fig.~\ref{Fig:oscillations} the early-time
behavior of the longitudinal component of the 
DM velocity (see Eq.~(\ref{pertbs}) for the definition) 
and the DM density contrast;
the latter is normalized to the density contrast  
in $\Lambda$CDM. One
observes that the oscillations are rather weak in the velocity
and are almost completely washed out in the density
contrast.  
Besides,  for the velocity they are perceptible only
in very short modes that have by now entered into the non-linear
regime. Thus it is unclear whether the oscillations can have an impact on any
observable.

We now consider the effects 
of different sets of parameters on the spectra of perturbations at the present moment of time.
 The different choices are listed in
Table~\ref{tabul} together with the corresponding values of the
screening scales $k_{Y,0}$ and $k_{Y,eq}$ (see Eqs.~(\ref{ky1}),
(\ref{kyeq}) for definitions).  All parameter choices are
consistent with the gravitational 
tests  described in Sec.~\ref{sec:aether}.
The initial spectrum is taken to be
flat with the same normalization in all cases.
\begin{table}[!h]
\begin{center}
\begin{tabular}{|c|c|c|c|c|c|c|}
\hline
  & $\alpha$         & $\beta$   & $\lambda$ & $Y$  &$k_{Y,0}$ (h
  Mpc$^{-1}$) 
 &$k_{Y,eq}$ (h Mpc$^{-1}$)  \\\hline
a & $2\cdot 10^{-2}$ & $10^{-2}$ & $10^{-2}$ & $0.2$
&$9.2\cdot 10^{-4}$ & $6.5\cdot 10^{-2}$ \\\hline
b & $2\cdot 10^{-4}$ & $10^{-4}$ & $10^{-4}$ & $0.2$   
&$9.1\cdot 10^{-3}$ & $0.65$ \\\hline
c & $2\cdot 10^{-4}$ & $10^{-4}$ & $10^{-4}$ & $0.02$
&$2.6\cdot 10^{-3}$  & $0.18$ \\\hline 
d & $10^{-7}$ & $0$ & $10^{-7}$ & $0.2$ &
$0.41$ & $29$ \\\hline 
\end{tabular}
\end{center}
\caption{The values of the parameters used in numerical simulations.
\label{tabul}}
\end{table}

\begin{figure}[!tb]
\begin{center}
\includegraphics[width=0.45\textwidth]{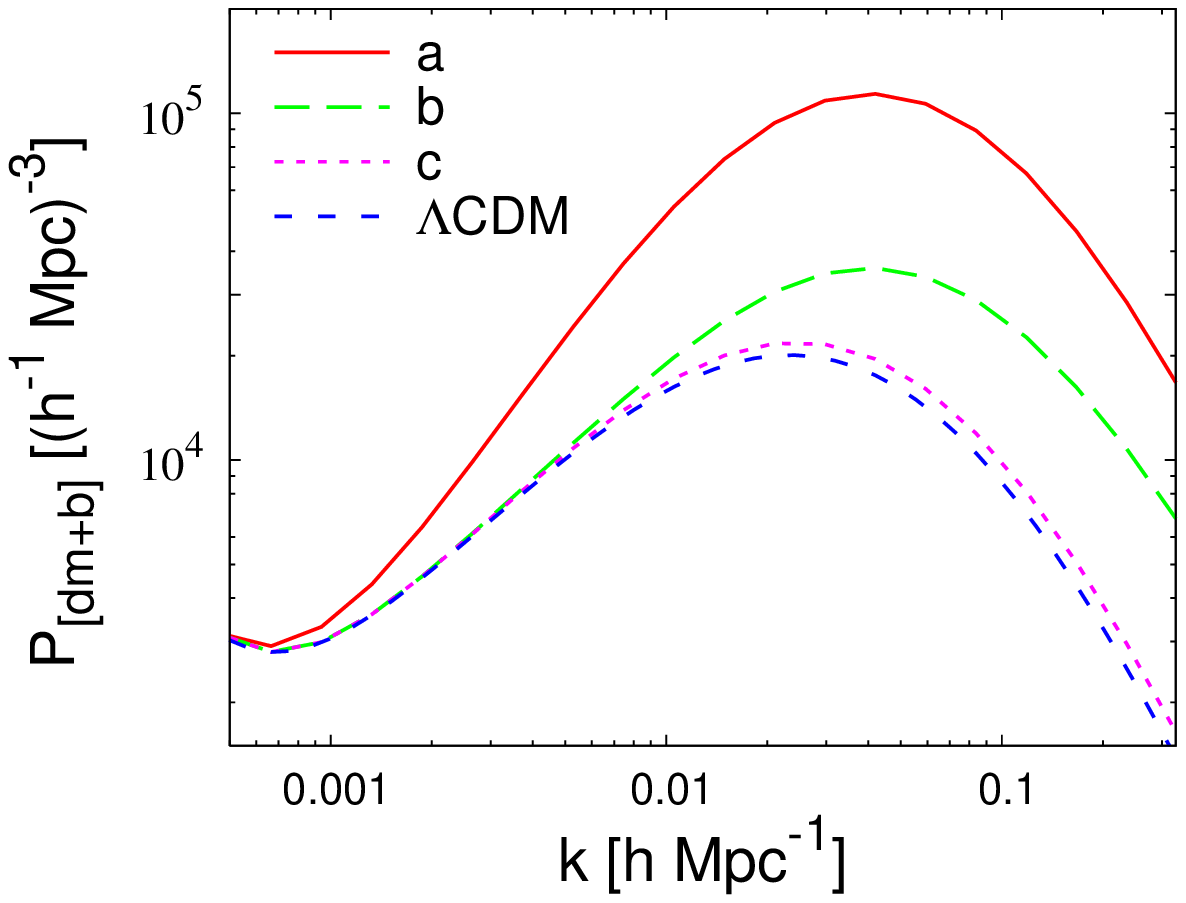}\qquad\quad
\includegraphics[width=0.45\textwidth]{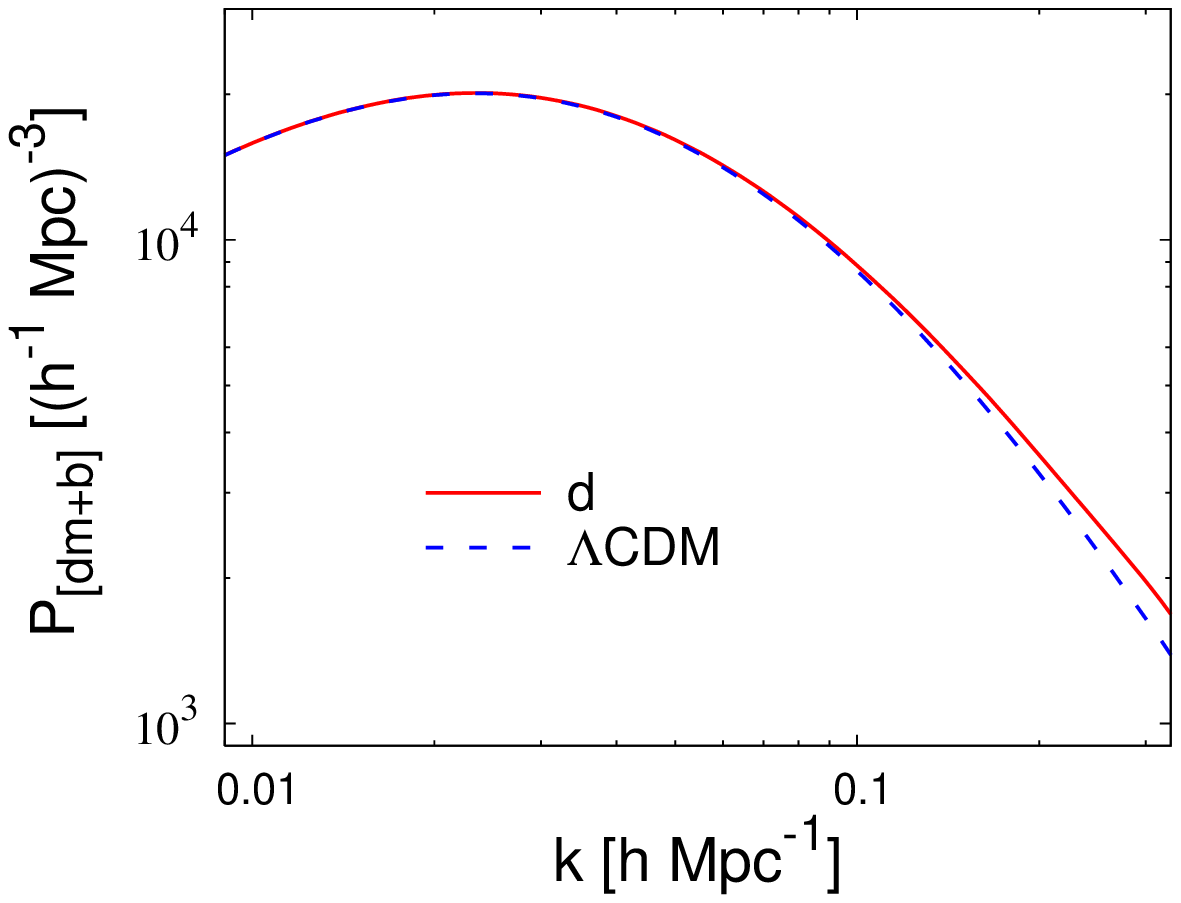}
\caption{Matter power spectrum for several values of the parameters
  listed in Table~\ref{tabul}. The case of $\Lambda$CDM is
  shown for comparison.
\label{Fig:5}
}
\end{center}
\end{figure} 
The comparison between the matter power spectrum in the LV models and in
$\Lambda$CDM is shown in Fig.~\ref{Fig:5}. The left panel shows
the cases when the present screening momentum $k_{Y,0}$ is 
lower than $k_{max}$ --- the position of the power spectrum maximum. We
clearly see the change in the slope of the spectrum in the interval
$k_{Y,0}<k<k_{Y,eq}$ accompanied by the shift of the position of the
maximum. The effect is significant for values of the parameter
$Y$ as low as a few per cent,
which suggests that these values can be tested observationally. The
right panel shows the situation when $k_{Y,0}$ is larger than
$k_{max}$, corresponding to very small values of the khronon
parameters $\alpha,\beta,\lambda$ and relatively strong LV in DM, see
Table~\ref{tabul}. The position of the maximum does not move in this
case but the change in the slope is still visible.   

\begin{figure}[!tb]
\begin{center}
\includegraphics[width=0.6\textwidth]{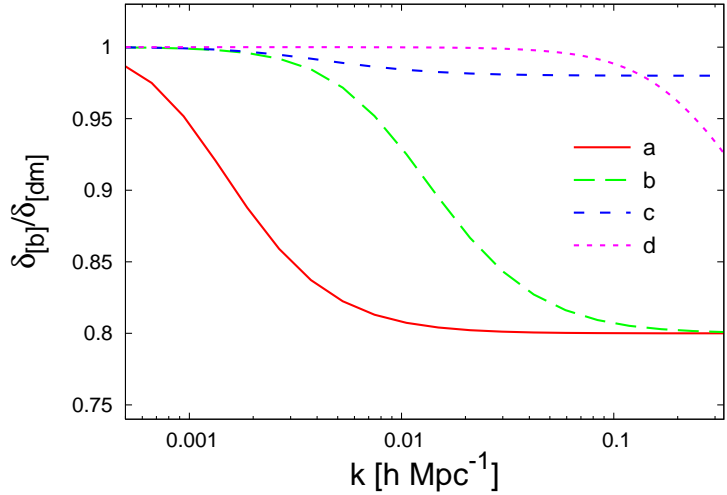}
\caption{Ratio between the amplitudes of perturbations in the baryonic
  and dark matter components. The curves correspond to the parameters 
 listed in Table~\ref{tabul}. 
\label{Fig:7}
}
\end{center}
\end{figure} 
Figure \ref{Fig:7} shows the ratio between the amplitudes of
perturbations in the baryonic and DM components. As expected
from the analytic considerations of Sec.~\ref{sec:CosmoPert}, this
ratio drops from 1 at $k<k_{Y,0}$ to $(1-Y)$ at larger momenta
implying a scale dependent bias between  baryons and DM. 

\begin{figure}[!tb]
\begin{center}
\includegraphics[width=0.6\textwidth]{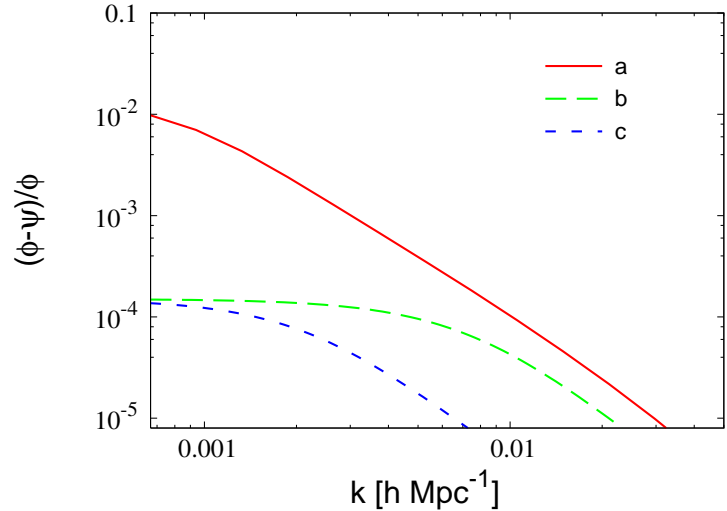}
\caption{Relative difference between the two scalar gravitational potentials
  for several choices of parameters listed in Table~\ref{tabul}. 
\label{Fig:6}
}
\end{center}
\end{figure} 
Finally, Fig.~\ref{Fig:6} presents the $k$-dependence of the 
``relative anisotropic stress'' --- the difference between the two
gravitational potentials $\phi$ and $\psi$ in the conformal Newton
gauge, normalized to $\phi$. We observe that at small momenta it has a
plateau with the magnitude set by the parameter $\beta$. The plateau
extends up to $k\approx k_{Y,0}$, beyond which the anisotropic stress
drops as the approximate power--law $k^{-2}$. All this is in agreement with the
analytic estimates of Sec.~\ref{sec:CosmoPert}. Note that the
persistence of the anisotropic stress up to relatively large
momentum, $k_{Y,0}\gg H_0$, is a peculiar signature of the present
model that contrasts with the more common situation where the anisotropic
stress quickly decays for subhorizon modes. 
Besides, the anisotropic stress is present at early times even for
adiabatic initial conditions. This distinguishes the present model 
from 
Lorentz-invariant models of modified gravity, such as $f(R)$
\cite{Sotiriou:2008rp} 
or DGP \cite{DGP} theories, where the
anisotropic stress is generated only at a recent epoch
\cite{Tsujikawa:2010zza}.

\section{Summary and discussion}\label{sec:conclusions}

In this paper we have studied the possibility to test 
the Lorentz invariance of  dark matter (DM)
with  cosmological observations. Our description is
based on the  Einstein-aether/ khronometric model, which 
provides an effective description of Lorentz invariance violation 
(LV) in gravitational relativistic theories. 
In those models, LV is encoded in a new field (aether) whose expectation
value determines a local preferred frame.
We considered DM as described by a pressureless fluid, 
and generalized its dynamics  to include the LV effects. 
Those effects amount to different couplings between
DM and the aether field. 
For cosmological perturbations in the linear regime 
all the LV effects in the DM sector can be summarized in a single 
parameter
$Y$.
This constant is to be added to the parameters of the
aether
sector (see Eqs.~(\ref{Saether}), (\ref{Kmnsr}) and (\ref{khpar})),
which are constrained by local tests of gravity. We
considered the Newtonian limit of the model and demonstrated that LV
implies modification of the inertial mass for small DM halos
thus leading to the violation of the equivalence principle. For large
halos this effect is screened by a variant of the chameleon mechanism.
Additionally, we pointed out the presence of a velocity-dependent
interaction; however, its 
role in the evolution of the Universe at large scales is negligible. 

The homogeneous expansion history of the Universe  for LV DM was
found to be exactly the same as in $\Lambda$CDM.
However, the evolution of linear cosmological perturbations 
presents 
three major effects permitting us to distinguish between the two
scenarios. 
The first effect is the accelerated
growth of inhomogeneities for the modes affected by the violation of the
equivalence principle. These are the modes that are short enough so
that the chameleon mechanism does not switch on.   
This effect eventually leads to the increase in the
slope of the matter power spectrum  with respect to $\Lambda$CDM
in a range of momenta.
The enhancement depends only on the parameter $Y$, whereas the
range of momenta is determined also by the aether parameters
(cf. Fig.~\ref{Fig:5}).  
The second effect is the appearance
of a new bias between the fluctuations of dark and
ordinary baryonic matter. Importantly, the bias exhibits
scale-dependence already at the linear level (cf. Fig.~\ref{Fig:7}).
Finally, the model predicts non-zero
anisotropic stress resulting in the difference between the
perturbations of the two gravitational potentials in the conformal
Newton gauge. While this effect is present already in the pure
Einstein-aether/khronometric theory, the novel feature introduced by
the aether -- DM coupling 
is the persistence of 
the anisotropic stress 
with time in a wide range of
momenta. As a consequence, 
its present-day power spectrum extends to 
wavelengths quite inside the current horizon (cf. Fig.~\ref{Fig:6}). 
 
A qualitative comparison between the predictions of the model and
those of $\Lambda$CDM suggests that the existing data (including
local tests of gravity) have
the potential to 
constrain deviations from Lorentz invariance in DM at the
level of a few per cent or even 
better, $Y\lesssim 0.01$. This limit
depends on the parameters of the aether sector and may be stronger or
weaker for certain regions in the parameter space. 
Detailed numerical simulations are required to set the precise bounds. This work is currently
in progress.    

It is worth comparing the signatures we found with the predictions of
Lorentz invariant
models where DM has unusual properties.
A large class
of models discussed in the literature 
\cite{FERMILAB-PUB-91-077-A,FERMILAB-PUB-92-008-A,Farrar:2003uw,Gubser:2004uh,
Farrar:2006tb,Bean:2008ac,Keselman:2009nx}
includes
 one or several light scalar fields mediating long-range interactions
between DM particles, referred to as fifth-forces.
Depending on the model, the
range of the fifth-force may or may not depend on the DM
density.
 The additional interaction implies violation of the
equivalence principle in the DM sector, similarly to our
model. 
As a consequence,  models with a  fifth-force
can also exhibit an accelerated growth of matter
perturbations, as well as  scale-dependent bias between DM 
and baryons. However, the resulting shape of the power spectrum
in these models is in general different from the one found in this
paper. The main reason is the different time dependence of the screening
scale for the fifth-force.
For example, in the
simplest case of a scalar with fixed mass $\mu$ the screening momentum
--- the analog of our $k_Y(\tau)$ --- is set by $\mu \,a(\tau)$ and thus
{\em grows} with time. This means that for a given mode the
fifth-force 
is important at early times and becomes screened later. On the
contrary, in
our case $k_{Y}(\tau)$ {\em decreases} as
$a(\tau)^{-1/2}$ and the modes enter into the regime of non-standard
evolution at late epochs. This leads to quite different shapes of
the transfer function and power spectrum in the two cases, compare 
Fig.~\ref{fig:pows} with
Fig.~4 of \cite{FERMILAB-PUB-92-008-A}.

One can envisage  more complicated 
fifth-force models where the squared mass of the scalar field is
proportional to the DM density. The screening scale in such a model will behave
exactly as $k_{Y}(\tau)$ mimicking the effect of LV on the power
spectrum. 
In this case one can try to distinguish between the two models by
examining the relation between the anomalous growth index of
perturbations and the bias factor. In the LV case this relation is set
by Eqs.~(\ref{NPpow}), (\ref{bias}), while in the fifth-force theories
the formulas are more complicated, cf. Eqs.~(3.11), (3.12) of
\cite{FERMILAB-PUB-92-008-A}. The origin of this discrepancy lies in
the different dynamics of the models. As discussed in
Sec.~\ref{sec:Newton}, the violation of
equivalence principle in our case is mainly due to the change in the {\em
inertial} mass of the DM particles (Eq.~(\ref{1peqNew})), while their {\em
gravitational} mass remains the same (Eq.~(\ref{Neweq})). The
gravitational potential 
produced by DM is just the usual one. Combined with the
contribution of baryons it adds up to the total potential that 
appears as the same forcing term in the hydrodynamical equations both for
baryons and DM (Eq.~(\ref{1peqNew}) and the equivalent with
$Y=0$ for baryons).  
In other words, the bias between
DM and
baryons in the case of LV comes from the  
change of the inertial mass of DM.
On the other hand, 
the standard 
fifth-force appears as an extra forcing term for the DM sector, without a modification
of the inertial mass of DM. 
Thus the hydrodynamic equations for baryons and DM have different
forcing terms with a complicated relation between them.

Another feature of our model that distinguishes it from other fifth-force
theories is the presence of the anisotropic stress both at early and late times.
However, this effect is rather small and may be outside the reach of observations
(cf. Fig.~\ref{Fig:6}).

There are several interesting  questions to explore in the
physics of LV DM.  For large scale structure, it would be interesting
to go beyond linear theory and understand how  LV 
affects the cosmological dynamics of DM at the non-linear level. This would allow us
to extend our results to
scales shorter than $100$ Mpc, where new data can be used to
constrain the model. For this, one can resort to 
analytic methods in the lines of 
\cite{Amendola:2003wa,Scoccimarro:2009eu,Saracco:2009df,Brax:2012sy}
or, alternatively, use N-body simulations.
Another open direction is related to the finding
\cite{astro-ph/0608095} that the violation of equivalence principle by
DM has pronounced effects on the tidal disruption of satellite
galaxies. It seems promising to look for similar signatures at the
level of galaxies and galaxy clusters in the model considered in this
paper. One expects to encounter a rather rich dynamics, given that
the model contains velocity-dependent interactions and automatically
incorporates a chameleon mechanism that screens the deviations from
the standard physics for large mass concentrations. This kind of study
may lead to interesting bounds on the model,
complementary to those discussed in the present work.

\paragraph*{Acknowledgments}

We are grateful to
 Cedric Deffayet, Sergei Demidov, Sergey Du\-bov\-sky,
Dmitry Gorbunov, Eugene Lim, Alexander Panin,
Valery Rubakov, Roman Scoccimarro, Igor Tkachev, Alexey Toporensky, 
Shinji Tsujikawa and 
Andrea Wulzer for useful 
discussions. 
D.B. thanks CCPP of NYU for hospitality at the earlier stages of this work.
M.I. and S.S. are grateful to the Institute of Theoretical Physics
of EPFL for hospitality during the completion of this paper.
This work was supported in part by the Swiss National Science Foundation, grant
IZKOZ2 138892/1 of the International Short Visits Program (D.B.),
the Grants of the President of Russian Federation
NS-5590.2012.2 and MK-3344.2011.2~(M.I. and S.S.), the RFBR grants 
11-02-92108 (S.S.), 11-02-01528 (S.S.), 12-02-01203 (S.S.)
and by the Dynasty Foundation (M.I. and S.S.). 

\appendix

\section{Numerical procedure}\label{numerics}

The complete system of equations for the evolution of the cosmological
perturbations in the model consists of Einstein's equations
(\ref{Einsts}), equations for the khronon and 
DM (\ref{eqs_perturbch}) and the 
hydrodynamical equations for the ordinary matter (\ref{matt}). Not all
of these equations are independent. 
For the numerical procedure we choose two Einstein's equations (\ref{Einstijtrace}),
(\ref{Einstijtracefree}). To these we add Eqs.~(\ref{khronoeq}),
(\ref{darkmeq}) for the evolution of the khronon and DM
velocity. To close the system we have to evaluate the pressure term in
(\ref{Einstijtrace}). Only the radiation component contributes into
it. The two first-order equations (\ref{matt3}) governing its
evolution can be reduced to one of second order,
\[
\ddot\delta_{[\gamma]}-\frac{\Delta\delta_{[\gamma]}}{3}
-\frac{4\Delta\phi}{3}-4\ddot\psi=0\;,
\] 
where we have introduced the density contrast,
\[
\delta_{[\gamma]}\equiv \frac{\delta\rho_{[\gamma]}}{\bar\rho_{[\gamma]}}\;.
\]
Performing the Fourier decomposition, normalizing the
present scale factor to one, $a(\tau_0)=1$, and choosing the units such
that the present Hubble parameter is equal to one, $H_0=1$, we obtain
the final system of 
ordinary differential equations to be solved numerically:
\bseq
\label{numeqs}
\begin{align}
\label{eq1}
&\ddot{\psi}+\mathcal{H}(\dot{\phi}+2\dot{\psi})+(2\dot{\mathcal{H}}+\mathcal{H}^2)\phi
-\frac{\beta+\lambda}{2+\alpha \mathcal B} k^2(\dot{\chi}+2\mathcal{H}\chi)
-\frac{\Omega_\gamma}{2a^2}\delta_{[\gamma]}=0\;,\\
\label{eq1a}
&\phi-\psi-\beta(\dot\chi+2\mathcal{H}\chi)=0\;,\\
\label{eq2}
&\ddot{\chi}+2\mathcal{H}\dot{\chi}+{\cal C}^2 k^2\chi
+\bigg[\dot{\mathcal{H}}(1-\mathcal B)
+\mathcal{H}^2(1+\mathcal B)
+\frac{3(2+\alpha \mathcal B)Y\Omega_{dm}}{2\alpha\, a}\bigg]\chi \notag\\
&\hspace{4cm}-\frac{3(2+\alpha \mathcal B)Y\Omega_{dm}}{2\alpha\, a}v_{[dm]} 
-\dot{\phi}-\mathcal{H}(1+\mathcal B)\phi -\mathcal B\dot{\psi}=0\;,\\
\label{eq3}
&\dot{v}_{[dm]}+{\cal H}v_{[dm]}+\frac{Y}{1-Y}
(\dot{\chi}+\mathcal{H}\chi)-\frac{\phi}{1-Y}=0 \;,\\
\label{eq4}
&\ddot{\delta}_{[\gamma]}+\frac{k^2}{3}\delta_{[\gamma]}+\frac{4k^2}{3}\phi
-4\ddot{\psi}=0\;.
\end{align}
\eseq

A subtle point is the proper choice of the initial conditions for the system
(\ref{numeqs}). These are fixed deep inside the radiation-domination stage
when the modes are superhorizon. We consider the initial conditions
corresponding to the adiabatic mode. The latter is regular at 
$\tau\to 0$. Thus we
write for small $\tau$:
\bseq
\label{expans}
\begin{gather}
\phi=\phi^{(0)}+\phi^{(1)}\tau\;,~~~\psi=\psi^{(0)}+\psi^{(1)}\tau\;,~~~
\delta_{[\gamma]}=\delta_{[\gamma]}^{(0)}+\delta_{[\gamma]}^{(1)}\tau\;,\\
\chi=\chi^{(1)}\tau+\chi^{(2)}\frac{\tau^2}{2}\;,~~~~
v_{[dm]}=v^{(1)}_{[dm]}\tau\;.
\end{gather}
\eseq
Expanding Eqs.~(\ref{numeqs}) at $\tau\to 0$ we obtain the
relations:
\bseq
\label{ini}
\begin{gather}
\label{inileading}
\delta_{[\gamma]}^{(0)}=-2\phi^{(0)}\;,\quad
\chi^{(1)}=v_{[dm]}^{(1)}=\frac{\phi^{(0)}}{2}\;,\quad
\psi^{(0)}=\bigg(1-\frac{3\beta}{2}\bigg)\phi^{(0)}\;,\\
\delta_{[\gamma]}^{(1)}=4\psi^{(1)}\;,\qquad
\phi^{(1)}-\psi^{(1)}-2\beta\chi^{(2)}
-\frac{\beta(\Omega_{dm}+\Omega_b)}{4\sqrt{\Omega_\gamma}}\phi^{(0)}=0\;,\\
(2+\mathcal B)\phi^{(1)}+\mathcal B\psi^{(1)}-(3+{\cal B})\chi^{(2)}
-\frac{\Omega_{dm}+\Omega_b}{4\sqrt{\Omega_\gamma}}\phi^{(0)}=0\;,
\end{gather}
\eseq
where we have used the expansion of the scale factor at the 
radiation-domination epoch including the subleading order 
\[
a=\sqrt{\Omega_{\gamma}}\tau+\frac{\Omega_{dm}+\Omega_b}{4}\tau^2 \;.
\] 
Note that Eqs.~(\ref{inileading}) describing the leading form of the
adiabatic mode agree with those of Sec.~\ref{ssec:5.1}.
Additionally, the initial data must satisfy   
the constraint following
from the $(00)$ Einstein's equation 
(\ref{Einst00})\footnote{It is straightforward to check that the
 $(0i)$ equation  
(\ref{Einst0i}) does not produce any new constraints.}. 
This gives,
\[
3\psi^{(1)}+\phi^{(1)}+
\frac{\Omega_{dm}+\Omega_b}{\sqrt{\Omega_\gamma}}\phi^{(0)}
+\frac{\Omega_{dm}}{2\sqrt{\Omega_\gamma}}\delta_{[dm]}^{(0)}
+\frac{\Omega_b}{2\sqrt{\Omega_\gamma}}\delta_{[b]}^{(0)}=0\;,
\]
where $\delta_{[dm]}^{(0)}$, $\delta_{[b]}^{(0)}$ are the constant
terms in the expansion of the DM and baryon density contrasts.
For the 
adiabatic mode we have:
\[
\delta_{[dm]}^{(0)}=\delta_{[b]}^{(0)}=\frac{3}{4}\delta_{[\gamma]}^{(0)}\;.
\]
Using this relation we obtain for the coefficients of the subleading terms in
(\ref{expans}),
\begin{gather}
\phi^{(1)}=\psi^{(1)}=\frac{\delta_{[\gamma]}^{(1)}}{4}=\frac{\chi^{(2)}}{2}=
-\frac{\Omega_{dm}+\Omega_b}{16\sqrt{\Omega_\gamma}}\phi^{(0)}\;.
\end{gather}
In this way all initial conditions for the system (\ref{numeqs}) are
fixed in terms of the overall amplitude $\phi^{(0)}$. 

Once the quantities entering the system (\ref{numeqs}) are computed, we
find the DM and baryon density contrasts by integrating the
equations
\bseq
\begin{gather}
\dot\delta_{[dm]}+k^2v_{[dm]}-3\dot\psi=0\;,\\
\dot v_{[b]}+{\cal H} v_{[b]}-\phi=0~,\quad
\dot\delta_{[b]}+k^2v_{[b]}-3\dot\psi=0
\end{gather} 
\eseq
with the initial conditions
\[
v_{[b]}=\frac{\phi^{(0)}}{2}\tau~,\qquad 
\delta_{[dm]}=\delta_{[b]}=-\frac{3}{2}\phi^{(0)}\;.
\]

By similar reasoning we also obtain the equations and initial conditions
for the case of $\Lambda$CDM. As we are interested in comparing the
evolution of perturbations
in the proposed model and $\Lambda$CDM we can
choose arbitrary normalization for $\phi^{(0)}$, the only requirement
being that this normalization is the same for the computations in both
models. In practice we take $\phi^{(0)}=1$ to compute the transfer
functions which we multiply by a
flat initial spectrum when appropriate.


\begin{thebibliography}{99}
\bibitem{Kostelecky:2008ts} 
  V.~A.~Kostelecky and N.~Russell,
  Rev.\ Mod.\ Phys.\  {\bf 83}, 11 (2011)
  [arXiv:0801.0287 [hep-ph]].

\bibitem{Weinberg:1965rz} 
  S.~Weinberg,
  Phys.\ Rev.\  {\bf 138}, B988 (1965).

\bibitem{Deser:1969wk} 
  S.~Deser,
  Gen.\ Rel.\ Grav.\  {\bf 1}, 9 (1970)
  [gr-qc/0411023].

\bibitem{Wald:1986bj} 
  R.~M.~Wald,
  Phys.\ Rev.\ D {\bf 33}, 3613 (1986).

\bibitem{Blas:2007pp} 
  D.~Blas,
  J.\ Phys.\ A A {\bf 40}, 6965 (2007)
  [hep-th/0701049].

\bibitem{Horava:2009uw} 
  P.~Horava,
  Phys.\ Rev.\ D {\bf 79}, 084008 (2009)
  [arXiv:0901.3775 [hep-th]].

\bibitem{Blas:2009ck}
  D.~Blas, O.~Pujolas and S.~Sibiryakov,
  Phys.\ Lett.\  B {\bf 688}, 350 (2010)
  [arXiv:0912.0550 [hep-th]].

\bibitem{Mattingly:2005re} 
  D.~Mattingly,
  Living Rev.\ Rel.\  {\bf 8}, 5 (2005)
  [gr-qc/0502097].

\bibitem{Jacobson:2005bg} 
  T.~Jacobson, S.~Liberati and D.~Mattingly,
  Annals Phys.\  {\bf 321}, 150 (2006)
  [astro-ph/0505267].

\bibitem{Blas:2009yd} 
  D.~Blas, O.~Pujolas and S.~Sibiryakov,
  JHEP {\bf 0910}, 029 (2009)
  [arXiv:0906.3046 [hep-th]].
  
\bibitem{Jacobson:2000xp}
  T.~Jacobson and D.~Mattingly,
  Phys.\ Rev.\  D {\bf 64}, 024028 (2001)
  [arXiv:gr-qc/0007031].

\bibitem{Jacobson:2008aj}
  T.~Jacobson,
  PoS {\bf QG-PH}, 020 (2007)
  [arXiv:0801.1547 [gr-qc]].

\bibitem{Withers:2009qg} 
  B.~Withers,
  Class.\ Quant.\ Grav.\  {\bf 26}, 225009 (2009)
  [arXiv:0905.2446 [gr-qc]].

\bibitem{Blas:2010hb}
  D.~Blas, O.~Pujolas, S.~Sibiryakov,
  JHEP {\bf 1104}, 018 (2011).
  [arXiv:1007.3503 [hep-th]].

\bibitem{Jacobson:2010mx}
  T.~Jacobson,
  Phys.\ Rev.\  D {\bf 81}, 101502 (2010)
  [Erratum-ibid.\  D {\bf 82}, 129901 (2010)]
  [arXiv:1001.4823 [hep-th]].

\bibitem{Blas:2009qj}
D.~Blas, O.~Pujolas and S.~Sibiryakov,
Phys.\ Rev.\ Lett.\ {\bf 104} (2010) 181302
[arXiv:0909.3525 [hep-th]].

\bibitem{Mukohyama:2010xz} 
  S.~Mukohyama,
  Class.\ Quant.\ Grav.\  {\bf 27}, 223101 (2010)
  [arXiv:1007.5199 [hep-th]].

\bibitem{Blas:2011zd}
  D.~Blas and H.~Sanctuary,
  Phys.\ Rev.\ D {\bf 84} (2011) 064004
  [arXiv:1105.5149 [gr-qc]].
  
\bibitem{GrootNibbelink:2004za}
  S.~Groot Nibbelink and M.~Pospelov,
  Phys.\ Rev.\ Lett.\  {\bf 94}, 081601 (2005)
  [arXiv:hep-ph/0404271].
  P.~A.~Bolokhov, S.~G.~Nibbelink and M.~Pospelov,
  Phys.\ Rev.\  D {\bf 72}, 015013 (2005)
  [arXiv:hep-ph/0505029].

\bibitem{Pujolas:2011sk} 
  O.~Pujolas and S.~Sibiryakov,
  JHEP {\bf 1201}, 062 (2012)
  [arXiv:1109.4495 [hep-th]].

\bibitem{Pospelov:2010mp} 
  M.~Pospelov and Y.~Shang,
  Phys.\ Rev.\ D {\bf 85}, 105001 (2012)
  [arXiv:1010.5249 [hep-th]].

\bibitem{Blas:2011en}
  D.~Blas, S.~Sibiryakov,
  JCAP {\bf 1107}, 026 (2011)
  [arXiv:1104.3579 [hep-th]].

\bibitem{Duffy:2009ig} 
  L.~D.~Duffy and K.~van Bibber,
  New J.\ Phys.\  {\bf 11}, 105008 (2009)
  [arXiv:0904.3346 [hep-ph]].

\bibitem{FERMILAB-PUB-91-077-A}
  J.~A.~Frieman and B.~-A.~Gradwohl,
  Phys.\ Rev.\ Lett.\ \ {\bf 67} (1991) 2926.

\bibitem{FERMILAB-PUB-92-008-A}
B.~-A.~Gradwohl and J.~A.~Frieman,
  Astrophys.\ J.\ \ {\bf 398} (1992) 407.

\bibitem{Bovy:2008gh} 
  J.~Bovy and G.~R.~Farrar,
  Phys.\ Rev.\ Lett.\  {\bf 102}, 101301 (2009)
  [arXiv:0807.3060 [hep-ph]].

\bibitem{Carroll:2008ub} 
  S.~M.~Carroll, S.~Mantry, M.~J.~Ramsey-Musolf and C.~W.~Stubbs,
  Phys.\ Rev.\ Lett.\  {\bf 103}, 011301 (2009)
  [arXiv:0807.4363 [hep-ph]];
  S.~M.~Carroll, S.~Mantry and M.~J.~Ramsey-Musolf,
  Phys.\ Rev.\ D {\bf 81}, 063507 (2010)
  [arXiv:0902.4461 [hep-ph]].

\bibitem{Elliott:2005va} 
  J.~W.~Elliott, G.~D.~Moore and H.~Stoica,
  JHEP {\bf 0508}, 066 (2005)
  [hep-ph/0505211].

\bibitem{Dubovsky:2005xd}
  S.~Dubovsky, T.~Gregoire, A.~Nicolis and R.~Rattazzi,
  JHEP {\bf 0603}, 025 (2006)
  [arXiv:hep-th/0512260].

\bibitem{Adams:2006sv} 
  A.~Adams, N.~Arkani-Hamed, S.~Dubovsky, A.~Nicolis and R.~Rattazzi,
  JHEP {\bf 0610}, 014 (2006)
  [hep-th/0602178].

\bibitem{Blas:2011ni}
D.~Blas and S.~Sibiryakov,
Phys.\ Rev.\ D {\bf 84} (2011) 124043
[arXiv:1110.2195 [hep-th]].

\bibitem{Carroll:2004ai}
  S.~M.~Carroll and E.~A.~Lim,
  Phys.\ Rev.\  D {\bf 70}, 123525 (2004)
  [arXiv:hep-th/0407149].

\bibitem{Foster:2005dk}
  B.~Z.~Foster and T.~Jacobson,
  Phys.\ Rev.\ D {\bf 73} (2006) 064015
  [gr-qc/0509083].

\bibitem{Will:2005va}
  C.~M.~Will,
  Living Rev.\ Rel.\  {\bf 9}, 3 (2005)
  [arXiv:gr-qc/0510072].

\bibitem{Will:Book}
  C.~M.~Will,
  ``Theory and experiment in gravitational physics,''
{\it  Cambridge, UK: Univ. Pr. (1993), 380 p.}

\bibitem{Foster:2006az} 
  B.~Z.~Foster,
  Phys.\ Rev.\ D {\bf 73}, 104012 (2006)
  [Erratum-ibid.\ D {\bf 75}, 129904 (2007)]
  [gr-qc/0602004]; 
  Phys.\ Rev.\ D {\bf 76}, 084033 (2007)
  [arXiv:0706.0704 [gr-qc]].

\bibitem{Liberati:2012th} 
  S.~Liberati and D.~Mattingly,
  ``Lorentz breaking effective field theory models for matter and gravity: theory and observational constraints,''
  arXiv:1208.1071 [gr-qc].

\bibitem{Coleman:1998ti} 
  S.~R.~Coleman and S.~L.~Glashow,
  Phys.\ Rev.\ D {\bf 59}, 116008 (1999)
  [hep-ph/9812418].

\bibitem{aetheralignment}
  I.~Carruthers and T.~Jacobson,
  Phys.\ Rev.\ D {\bf 83}, 024034 (2011)
  [arXiv:1011.6466 [gr-qc]].

\bibitem{Khoury:2003aq} 
  J.~Khoury and A.~Weltman,
  Phys.\ Rev.\ Lett.\  {\bf 93}, 171104 (2004)
  [astro-ph/0309300].

\bibitem{Andersson:2006nr}
  N.~Andersson and G.~L.~Comer,
  Living Rev.\ Rel.\  {\bf 10} (2005) 1
  [gr-qc/0605010].

\bibitem{ArmendarizPicon:2010rs}
C.~Armendariz-Picon, N.~F.~Sierra and J.~Garriga,
JCAP {\bf 1007} (2010) 010
[arXiv:1003.1283 [astro-ph.CO]].

\bibitem{Ma:1995ey} 
  C.~-P.~Ma and E.~Bertschinger,
  Astrophys.\ J.\  {\bf 455}, 7 (1995)
  [astro-ph/9506072].

\bibitem{Sotiriou:2008rp} 
  T.~P.~Sotiriou and V.~Faraoni,
  Rev.\ Mod.\ Phys.\  {\bf 82}, 451 (2010)
  [arXiv:0805.1726 [gr-qc]].

\bibitem{DGP} 
  G.~R.~Dvali, G.~Gabadadze and M.~Porrati,
  Phys.\ Lett.\ B {\bf 485}, 208 (2000)
  [hep-th/0005016].

\bibitem{Tsujikawa:2010zza} 
  S.~Tsujikawa,
  Lect.\ Notes Phys.\  {\bf 800}, 99 (2010)
  [arXiv:1101.0191 [gr-qc]].

\bibitem{Farrar:2003uw} 
  G.~R.~Farrar and P.~J.~E.~Peebles,
  Astrophys.\ J.\  {\bf 604}, 1 (2004)
  [astro-ph/0307316].

\bibitem{Gubser:2004uh} 
  S.~S.~Gubser and P.~J.~E.~Peebles,
  Phys.\ Rev.\ D {\bf 70}, 123510 (2004)
  [hep-th/0402225]; 
  Phys.\ Rev.\ D {\bf 70}, 123511 (2004)
  [hep-th/0407097].

\bibitem{Farrar:2006tb} 
  G.~R.~Farrar and R.~A.~Rosen,
  Phys.\ Rev.\ Lett.\  {\bf 98}, 171302 (2007)
  [astro-ph/0610298].


\bibitem{Bean:2008ac} 
  R.~Bean, E.~E.~Flanagan, I.~Laszlo and M.~Trodden,
  Phys.\ Rev.\ D {\bf 78}, 123514 (2008)
  [arXiv:0808.1105 [astro-ph]].

\bibitem{Keselman:2009nx}
J.~A.~Keselman, A.~Nusser and P.~J.~E.~Peebles,
Phys.\ Rev.\ D {\bf 81} (2010) 063521
[arXiv:0912.4177 [astro-ph.CO]].

\bibitem{Amendola:2003wa} 
  L.~Amendola,
  Phys.\ Rev.\ D {\bf 69}, 103524 (2004)
  [astro-ph/0311175].

\bibitem{Scoccimarro:2009eu}
  R.~Scoccimarro,
  Phys.\ Rev.\ D {\bf 80} (2009) 104006
  [arXiv:0906.4545 [astro-ph.CO]].

\bibitem{Saracco:2009df} 
  F.~Saracco, M.~Pietroni, N.~Tetradis, V.~Pettorino and G.~Robbers,
  Phys.\ Rev.\ D {\bf 82}, 023528 (2010)
  [arXiv:0911.5396 [astro-ph.CO]].

\bibitem{Brax:2012sy} 
  P.~Brax and P.~Valageas,
  ``Structure Formation in Modified Gravity Scenarios,''
  arXiv:1205.6583 [astro-ph.CO].

\bibitem{astro-ph/0608095}
  M.~Kesden and M.~Kamionkowski,
  Phys.\ Rev.\ D\ {\bf 74} (2006) 083007
  [astro-ph/0608095].

\end{thebibliography}
\end{document}